\documentclass[fleqn,usenatbib]{mnras}

\usepackage{newtxtext,newtxmath}

\usepackage[T1]{fontenc}

\DeclareRobustCommand{\VAN}[3]{#2}
\let\VANthebibliography\thebibliography
\def\thebibliography{\DeclareRobustCommand{\VAN}[3]{##3}\VANthebibliography}

\usepackage{booktabs, multirow} 
\usepackage{soul}
\usepackage[table]{xcolor} 
\usepackage{changepage,threeparttable} 
\usepackage{adjustbox}
\usepackage[figuresright]{rotating}

\usepackage{graphicx}	
\usepackage{amsmath}	
\usepackage{amssymb}	
\usepackage{xcolor}
\usepackage{float}
\restylefloat{figure}
\usepackage{caption}
\usepackage[none]{hyphenat}

\newcommand{\psrchive}{\texttt{PSRCHIVE}}
\newcommand{\pp}{\texttt{PulsePortraiture}}

\newcommand{\tempotwo}{\texttt{tempo2}}
\newcommand{\libstempo}{\texttt{libstempo}}
\newcommand{\pinta}{\texttt{pinta}}
\newcommand{\psrfits}{\texttt{PSRFITS}}
\newcommand{\dmcalc}{\texttt{DMcalc}}

\title{Multi-band Extension of the Wideband Timing Technique}

\author[]{Avinash Kumar Paladi$^{1,2}$\thanks{E-mail: avinashkumarpaladi@gmail.com},
 Churchil Dwivedi$^{2}$, Prerna Rana$^{3}$, Nobleson K.$^{4}$, Abhimanyu Susobhanan$^{5}$, 
 \newauthor Bhal Chandra Joshi$^{6,7}$, Pratik Tarafdar$^{8}$, Debabrata Deb$^{8}$, Swetha Arumugam$^{10}$, A. Gopakumar$^{3}$, 
 \newauthor M. A. Krishnakumar$^{11,12}$, Neelam Dhanda Batra$^{13}$, Jyotijwal Debnath$^{8,9}$, Fazal Kareem$^{14,15}$, 
 \newauthor Paramasivan Arumugam$^{7}$, Manjari Bagchi$^{8,9}$, Adarsh Bathula$^{16}$, Subhajit Dandapat$^{3}$, Shantanu Desai$^{17}$,  
 \newauthor Yashwant Gupta$^{6}$, Shinnosuke Hisano$^{18}$, Divyansh Kharbanda$^{17}$, Tomonosuke Kikunaga$^{18}$, Neel Kolhe$^{19}$, 
 \newauthor Yogesh Maan$^{6}$, P. K. Manoharan$^{20}$, Jaikhomba Singha$^{7}$, Aman Srivastava$^{17}$, Mayuresh Surnis$^{21}$, 
 \newauthor Keitaro Takahashi$^{22,23}$
\\ \\
$^{1}$ Joint Astronomy Programme, Indian Institute of Science, Bengaluru, Karnataka, 560012, India \\
$^{2}$ Department of Earth and Space Sciences, Indian Institute of Space Science and Technology, Valiamala, Thiruvananthapuram 695547, Kerala, India \\
$^{3}$ Department of Astronomy and Astrophysics, Tata Institute of Fundamental Research, Homi Bhabha Road, Navy Nagar, Colaba, Mumbai 400005, India \\
$^{4}$ Department of Physics, BITS Pilani Hyderabad Campus, Hyderabad 500078, Telangana, India \\ 
$^{5}$ Center for Gravitation Cosmology and Astrophysics, University of Wisconsin-Milwaukee, Milwaukee, WI 53211, USA \\
$^{6}$ National Centre for Radio Astrophysics, Pune University Campus, Pune 411007, India \\
$^{7}$ Department of Physics, Indian Institute of Technology Roorkee, Roorkee-247667, India \\
$^{8}$ The Institute of Mathematical Sciences, C. I. T. Campus, Taramani, Chennai 600113, India \\ 
$^{9}$ Homi Bhabha National Institute, Training School Complex, Anushakti Nagar, Mumbai 400094, India \\
$^{10}$ Department of Electrical Engineering, IIT Hyderabad, Kandi, Telangana 502284, India \\ 
$^{11}$ Max-Planck-Institut f\"ur Radioastronomie, Auf dem H\"ugel 69, 53121 Bonn, Germany \\ 
$^{12}$ Fakult\"at f\"ur Physik, Universit\"at Bielefeld, Postfach 100131, 33501 Bielefeld, Germany \\ 
$^{13}$ Department of Physics and Astrophysics, University of Delhi, Delhi \\ 
$^{14}$ Department of Physical Sciences, Indian Institute of Science Education and Research Kolkata, West Bengal, 741246, India \\ 
$^{15}$ Center of Excellence in Space Sciences India, Indian Institute of Science Education and Research Kolkata, West Bengal, 741246, India \\ 
$^{16}$ Indian Institute of Science Education and Research, Mohali - 140306, Punjab, India \\ 
$^{17}$ Department of Physics, IIT Hyderabad, Kandi, Telangana 502284, India \\ 
$^{18}$ Kumamoto University, Graduate School of Science and Technology, Kumamoto, 860-8555, Japan \\ 
$^{19}$ Department of Physics, St. Xavier’s College (Autonomous), Mumbai 400001, Maharashtra, India \\ 
$^{20}$ Arecibo Observatory, University of Central Florida, Arecibo 00612, USA \\ 
$^{21}$ Department of Physics, IISER Bhopal, Bhauri Bypass Road, Bhopal 462066, India \\ 
$^{22}$ Faculty of Advanced Science and Technology, Kumamoto University, 2-39-1 Kurokami, Kumamoto 860-8555, Japan \\ 
$^{23}$ International Research Organization for Advanced Science and Technology, Kumamoto University, 2-39-1 Kurokami, Kumamoto 860-8555, Japan
}

\date{Accepted XXX. Received YYY; in original form ZZZ}

\pubyear{2023}

\begin{document}
\label{firstpage}
\pagerange{\pageref{firstpage}--\pageref{lastpage}}
\maketitle

\begin{abstract}
The wideband timing technique enables the high-precision simultaneous estimation of pulsar Times of Arrival (ToAs) and Dispersion Measures (DMs) while effectively modeling frequency-dependent profile evolution. 
We present two novel independent methods that extend the standard wideband technique to handle simultaneous multi-band pulsar data incorporating profile evolution over a larger frequency span to estimate DMs and ToAs with enhanced precision.
We implement the wideband likelihood using the \libstempo{} python interface to perform wideband timing in the \tempotwo{} framework.
We present the application of these techniques to the dataset of fourteen millisecond pulsars observed simultaneously in Band 3 ($300-500$ MHz) and Band 5 ($1260-1460$ MHz) of the upgraded Giant Metrewave Radio Telescope (uGMRT) with a large band gap of 760 MHz as a part of the Indian Pulsar Timing Array (InPTA) campaign. 
We achieve increased ToA and DM precision and sub-microsecond root mean square post-fit timing residuals by combining simultaneous multi-band pulsar observations done in non-contiguous bands \textit{for the first time} using our novel techniques.
\end{abstract}




\section{Introduction }
Pulsars are rotating neutron stars emitting broadband electromagnetic radiation that is observed as periodic pulses.
The rotation of a pulsar can be tracked accurately by measuring the times of arrival (ToAs) of its pulses, and this technique is known as pulsar timing \citep{HobbsEdwardsManchester2006, EdwardsHobbsManchester2006}.
The pulsar signal is dispersed while propagating through the ionized interstellar medium (IISM) by an amount that is proportional to the integrated free electron column density along the line of sight, usually referred to as the dispersion measure (DM), and inversely proportional to the square of the observing frequency $\nu$ \citep{LorimerKramer2012}. 
Conventionally, the rough measurement of DM for a pulsar used to be done by splitting the data into multiple sub-bands and correcting for the DM induced delay for each sub-band and then adding the dispersed bands again \citep[][]{LorimerKramer2012}. In recent days, many sophisticated techniques have been proposed, which not only provides more accurate values of DM but also provide epoch to epoch variations of DM \citep[e.g.][]{AhujaGupta+2005}.

Pulsar timing has traditionally been done by splitting the data into multiple sub-bands with negligible dispersion smear and independently measuring the ToA from each sub-band, known as narrowband timing \citep{Taylor1992}. 
The improvement in telescope sensitivity, the advent of wideband receivers and backends \citep[e.g.][]{JohnstonSobey+2021,HobbsManchester+2020,ReddyKudale+2017,GuptaAjithkumar+2017}, and decades-long pulsar timing campaigns such as Pulsar Timing Arrays \citep[PTAs:][]{FosterBacker1990} have presented significant challenges to the narrowband approach.
These challenges include inadequate modeling of the pulse profile variability as a function of frequency, difficulty in correcting for interstellar scattering, and large data sizes.
The wideband timing technique seeks to address these issues by treating the pulse profile as a two-dimensional entity in frequency and pulse phase (usually referred to as a \textit{portrait}) and simultaneously measuring one ToA and one DM per observation using the full bandwidth \citep{PennucciDemorestRansom2014, Pennucci2019}.

PTA experiments, such as the Parkes Pulsar Timing Array \citep[PPTA:][]{Hobbs2013}, the European Pulsar Timing Array \citep[EPTA:][]{KramerChampion2013}, the North American Nanohertz Observatory for Gravitational Waves \citep[NANOGrav:][]{McLaughlin2013}, the Indian Pulsar Timing Array \citep[InPTA:][]{JoshiArumugasamy+2018}, the Chinese Pulsar Timing Array \citep[CPTA:][]{Lee2016}, MeerKat Pulsar Timing Array \citep[MPTA:][]{Miles+2023}, and the International Pulsar Timing Array \citep[IPTA:][]{HobbsArchibald+2010,PereraDeCesar+2019,VerbiestLentati+2016} consortium which combines the
data and resources from various PTAs, aim to detect nanohertz gravitational waves using an ensemble of millisecond pulsars (MSPs) as celestial clocks. Recent wideband timing studies across a wide range of observing frequencies have demonstrated significant improvements in ToA and DM measurement precision \citep{LiuDesvignes+2014,FonsecaCromartie+2021,NoblesonAgarwal+2022}, and PTAs and other high-precision pulsar timing campaigns are now increasingly adopting wideband techniques due to their advantages in dealing with broadband observations \citep[e.g.][]{AlamArzoumanian+2021,TarafdarNobleson+2022,CuryloPennucci+2022}.

The InPTA experiment complements the international PTA efforts by employing the unique features of the upgraded Giant Metrewave Radio Telescope \citep[uGMRT:][]{GuptaAjithkumar+2017}.
The high sensitivity of the uGMRT at low frequencies, combined with its ability to perform simultaneous multi-band observations, makes it an ideal instrument to characterize interstellar medium effects, which are stronger at low frequencies \citep{KrishnakumarManoharan+2021}.
The application of the wideband technique to the uGMRT data and the ToA and DM precision improvements therefrom were demonstrated in \citet{NoblesonAgarwal+2022}.
The recently published first data release of the InPTA \citep[InPTA DR1:][]{TarafdarNobleson+2022} built on the work of \citet{KrishnakumarManoharan+2021} and \citet{NoblesonAgarwal+2022}, has presented the results of narrowband and wideband timing of 14 pulsars observed over a time span of 3.5 years.
This work included some of the most precise DM measurements to date, estimated using both the narrowband and the wideband techniques. Recently the InPTA collaboration has completed Single Pulsar Noise Analysis on the DR1 pulsars using narrowband data \citep[][]{Srivastava+2023}.

The InPTA observes pulsars in two uGMRT bands, namely the Band 3 ($300-500$ MHz) and the Band 5 ($1260-1460$ MHz).
Although \citet{NoblesonAgarwal+2022} and \citet{TarafdarNobleson+2022} only used Band 3 data for the wideband timing, the DM precision achieved therein was comparable to the combined Band 3+5 narrowband DM estimates.
This raises the exciting possibility of attaining further improvements in DM precision by combining the two uGMRT bands in the wideband paradigm. 

In this work, we develop two novel methods to combine simultaneous observations of the same pulsar obtained at multiple bands using the wideband technique to obtain a single ToA and DM combination per epoch across these multiple bands. 
We then demonstrate the ToA and DM precision improvement achieved with these techniques using the InPTA observations of 14 MSPs, simultaneously observed at Band 3 and Band 5 seperated by a large band gap of about 760 MHz, which were selected for the InPTA first data release \citep{TarafdarNobleson+2022}.
These techniques provide significant improvements in the DM precision estimation that can be achieved using existing and future telescopes which can perform simultaneous or quasi-simultaneous multi-frequency observations such as the Square Kilometer Array (SKA) \citep{KramerStappers+2015,JanssenHobbs+2015}. For the timing analysis, we extend the traditional timing methodology to incorporate the wideband timing likelihood function (Appendix B of \citet{AlamArzoumanian+2021}) in \tempotwo{} using \libstempo{}. 


This paper is structured as follows. 
We begin by providing a brief overview of the wideband timing technique in subsection \ref{sec:wb-overview}.
In subsection \ref{sec:wb-combine-methods}, we present two novel independent methods for applying the wideband technique to two simultaneous band observations of a pulsar taken at different radio frequencies, which can be easily extended to multiple bands. We apply our methods to the case of PSR J1909$-$3744 and show the validation scheme and comparisons against each other as well as against the single band (Band 3) results in Section \ref{sec:validation}.
We present the application of our novel methods to the InPTA dataset of 14 MSPs in Section \ref{sec:inpta-data-appl}. We summarize our results in Section \ref{sec:summary} and discuss avenues for future improvements and extensions in Section \ref{sec:discussion}.
Our implementation of the wideband likelihood function using \tempotwo{} and \libstempo{} is briefly described in Appendix \ref{appB}.

\section{Multi-band extension of the Wideband Timing technique}

\subsection{Brief overview of the Wideband Technique}
\label{sec:wb-overview}
We begin by briefly summarising the wideband technique developed in \citet{PennucciDemorestRansom2014}, \citet{Pennucci2019}, and \citet{AlamArzoumanian+2021}.
The total intensity integrated pulse profile of a pulsar can be expressed as a two-dimensional object $D(\nu,\varphi)$ which is a function of the observing frequency $\nu$ and the pulse phase $\varphi$, and is referred to as a \textit{pulse portrait}.
Given a model for the observed portrait $P(\nu,\varphi)$, referred to as the \textit{template portrait} or the \textit{model portrait}, $D(\nu,\varphi)$ can be written as:
\begin{equation}
    D(\nu,\varphi) = B(\nu) + a(\nu) P(\nu,\varphi-\phi(\nu)) + N(\nu,\varphi)\,,
    \label{eq:wb-model}
\end{equation}
where $B(\nu)$ is the DC offset in each frequency channel, $a(\nu)$ is an amplitude that arises from the intrinsic power spectral density of the pulsar emission and interstellar scintillation and also depends on the receiver bandpass, and $N(\nu,\varphi)$ is an additive noise that is usually assumed to be Gaussian and uncorrelated in the absence of radio frequency interference (RFI).
In practice, $D(\nu,\varphi)$ and  $P(\nu,\varphi)$ are discretised in both $\nu$ and $\varphi$, i.e., $D_{nj}\equiv D(\nu_n,\varphi_j)$ such that $n$ denotes the  frequency channels and $j$ corresponds to the  phase bins.
The frequency dependence of the phase shift $\phi(\nu)$ arises primarily due to the interstellar dispersion and is given by:
\begin{equation}
    \phi(\nu) = \phi_{\text{ref}} + \frac{K\times \text{DM}}{P_s}\left(\frac{1}{\nu^2} - \frac{1}{\nu_\text{ref}^2}\right)\,,
    \label{eq:phase-shift}
\end{equation}
where $\phi_{\text{ref}}$ is the achromatic phase shift, $K$ is the Dispersion constant, $P_s$ is the apparent spin period of the pulsar at the epoch of observation, and $\nu_\text{ref}$ is a Barycentric reference frequency.
Given $\phi_{\text{ref}}$, the ToA can be computed as $t = t_f + {P_s\phi_{\text{ref}}}/{2\pi}$ where $t_f$ is the timestamp corresponding to a fiducial phase point in the data portrait\footnote{In practice, $t_f$ may be affected by instrumental delays such as those encountered in \citet{TarafdarNobleson+2022}, and one must correct for them.}.
$\phi_{\text{ref}}$ can be understood as the difference between the fiducial phases of the data portrait and the template portrait. 

Computing the discrete Fourier transform of equation \eqref{eq:wb-model} along the $\phi$ axis, applying the discrete Fourier shift theorem, and excluding the DC term, we have:
\begin{equation}
    \widetilde{D}_{nk} = a_n \widetilde{P}_{nk}\;e^{2\pi i k \phi_n} + \widetilde{N}_{nk}\,.
\end{equation}
where $\widetilde{D}_{nk}$ and $\widetilde{P}_{nk}$ denote the discrete Fourier transform of the data portrait and the template portrait respectively, and $\phi_n=\phi(\nu_n)$.
The quantities of interest $\phi_{\text{ref}}$ and $\text{DM}$ can then  be estimated by minimizing the weighted least-squares statistic:
\begin{equation}
    \chi^2(\phi_{\text{ref}},\text{DM},a_n) = \sum_{n,k}\frac{\left|\widetilde{D}_{nk} - a_n \widetilde{P}_{nk} e^{2\pi i k \phi_n} \right|^2}{{\sigma_n'}^2}\,,
\end{equation}
where ${\sigma_n'}^2$ denotes the noise variance of the Fourier coefficients $\widetilde{D}_{nk}$.
It turns out that $\chi^2(\phi_{\text{ref}},\text{DM},a_n)$ can be analytically minimised over the amplitudes $a_n$, and this leads to
\begin{equation}
    \chi^2(\phi_{\text{ref}},\text{DM}) = S - \sum_n \frac{C_n^2}{T_n}\,,
    \label{eq:wb-chisq}
\end{equation}
where 
\begin{subequations}
\begin{align}
    S &= \sum_{n,k} \frac{|\widetilde{D}_{nk}|^2}{{\sigma_n'}^2}\,,\\
    T_n &= \sum_{k} \frac{|\widetilde{P}_{nk}|^2}{{\sigma_n'}^2}\,,\\
    C_n &= \Re\left\{\sum_k \frac{\widetilde{D}_{nk}\widetilde{P}_{nk}^{*}\;e^{2\pi i k \phi_n}}{{\sigma_n'}^2}\right\}
    \,.
\end{align}
\label{eq:chisq_funcs}
\end{subequations}
Choosing $\nu_\text{ref}$ such that the covariance between $\phi_{\text{ref}}$ and DM vanishes (see the Appendix of \citet{PennucciDemorestRansom2014}), $\phi_{\text{ref}}$ and DM can be estimated by numerically minimizing $\chi^2(\phi_{\text{ref}},\text{DM})$.

The template portrait $P(\nu,\varphi)$ is usually obtained from a single high signal-to-noise ratio (S/N) portrait or an averaged portrait generated from many observations. The mean-subtracted template portrait is decomposed into many `eigenprofiles' using  principal component analysis (PCA). A smoothed template portrait is then reconstructed from a small number of significant eigenprofiles by spline-interpolating them \citep{Pennucci2019}. By linearly combining the $n_{\text{eig}}$ significant eigenprofiles $\hat{e}_i$ using the spline coefficients $B_i$ and adding it to the mean profile $\widetilde{p}$, a template profile $T(\nu)$ at any frequency $\nu$ can be created as
\begin{equation}
T(\nu) = \sum_{i=1}^{n_{\text{eig}}} B_i(\nu)\;\hat{e}_i + \widetilde{p} \,.
\label{eq:template}
\end{equation}

Note that the DMs estimated from the wideband technique are not derived from ToAs unlike in the narrowband case, but rather measured simultaneously with each ToA.
Therefore, DM measurements should be treated as data points on an equal footing with the ToAs while computing the likelihood function.
In the simple case of a pulsar with timing model parameters ($\boldsymbol{\theta}$), ToAs ($t_i$), timing residuals ($r_i$), ToA uncertainties ($\sigma_i$), DM measurements ($\text{DM}_i$), DM uncertainties ($\varsigma_i$), and DM model ($d(t)$),  the wideband log-likelihood can be written as
\begin{equation}
\ln L(\boldsymbol{r},\boldsymbol{\sigma},\boldsymbol{\text{DM}},\boldsymbol{\varsigma}|\boldsymbol{\theta}) = \ln L_0 - \frac{1}{2}\sum_i\left[\left(\frac{r_i}{\sigma_i}\right)^2  - \left(\frac{\text{DM}_i - d(t_i)}{\varsigma_i}\right)^2\right]\,,
\end{equation}
where the first term is a normalization term, the second term is the usual narrowband likelihood, and the third term is the likelihood function of the DM measurements.
A more general version of the above equation, applicable to more rigorous noise models can be found in Appendix B of \citet{AlamArzoumanian+2021}.

\subsection{Extending the Wideband technique for multiple bands}
\label{sec:wb-combine-methods}
The standard wideband timing technique, summarized in section \ref{sec:wb-overview}, has been applied to various single band observations across a wide range of observing frequencies \citep{FonsecaCromartie+2021,AlamArzoumanian+2021,NoblesonAgarwal+2022,TarafdarNobleson+2022,CuryloPennucci+2022}. In this section, we present and demonstrate two novel independent methods namely the \textit{Combined Portrait (CP)} method and \textit{Combined Chi-squared (CC)} method to combine simultaneous observations performed in two non-contiguous frequency bands within the paradigm of wideband technique to estimate a single ToA and DM combination per epoch covering the entire frequency range of these bands. It is straightforward to extend these techniques to multiple bands with simultaneous observations, which will be part of a future work.

\subsubsection{The Combined-portrait (CP) method}
\label{subsec:cp_description}
In this method, we begin by time-collapsing the frequency-resolved profiles obtained simultaneously in the two frequency bands using the \texttt{pam} command of \psrchive{} \citep{HotanvanStratenManchester2004}. We then combine the profiles in the two bands along the frequency axis using the \texttt{psradd} command of \psrchive{}. This requires both profiles to have the same number of phase bins; hence, we phase-collapse the higher-phase resolution profile to match the lower-phase resolution one using the \texttt{pam} command before appending them using \texttt{psradd}. The profiles of each frequency band are also collapsed in frequency to an appropriate number of sub-bands such that there is a reasonable signal-to-noise ratio (S/N) in each sub-band, and there are also enough sub-bands to obtain a 2-D template containing information of profile evolution across the band. Since the exact start time of the observation in each band may not be identical, the profiles are aligned by the \texttt{psradd} command by rotating them in phase using the pulsar ephemeris used for folding. For generating a noise-free template portrait, we use an epoch with high-S/N in both the bands. We first excise frequency channels with any residual RFI from both the  bands for the template epoch using the \texttt{pazi} command and then obtain a combined data profile using \texttt{psradd} covering the frequency of the two bands. Finally, a template portrait is generated from this combined data profile using the \texttt{ppalign} and \texttt{ppspline} modules of \pp{} \citep{PennucciDemorestRansom2014, Pennucci2019}. Here, the spline model is interpolated over the large frequency gap in between the two bands. For accurate modeling of the profile evolution across the two bands, we choose the required number of eigenprofiles and tolerance values for the template portrait using the procedure described in section 4.2 of \cite{TarafdarNobleson+2022}. A single wideband ToA and the corresponding DM for the combined observation of each epoch are then estimated using the \texttt{ppToAs} module of \pp{}.

\subsubsection{The Combined Chi-squared (CC) method} 
\label{subsec:cc_description}
In this method, we treat the data portraits and the corresponding templates for each band in their native phase resolution (without phase-collapsing or combining them along the frequency axis) and bandwidths. We use the time-collapsed data of two bands and partially collapse the frequency channels in each band to maintain a reasonable S/N in each sub-band. The noise-free templates are generated for each band separately using a high-S/N epoch after RFI excision. Here, there is no interpolation of spline model over the large frequency gap in between two bands, as both bands are treated separately. While generating the templates, we take care of the phase offset between multiple bands by rotating them appropriately. We estimate a single ToA and DM pair for multiple bands in each epoch by minimizing a combined chi-squared statistic defined as
\begin{equation}
    \chi^2(\phi_{\text{ref}},\text{DM}) = \sum_{b}\left\{S_b - \sum_n \frac{C_{bn}^2}{T_{bn}}\right\}\,,
    \label{eq:wb-chisq-comb}
\end{equation}
where the index $b$ labels the different bands, and $S_b$, $C_{bn}$ and $T_{bn}$ are defined by equations \eqref{eq:chisq_funcs} using the data portrait $D_{bnj}$ and the template portrait $P_{bnj}$ for each band $b$.
Since the timestamp $t_{fb}$ corresponding to the fiducial phase for different bands need not be the same, equation \eqref{eq:phase-shift} should be modified as follows:
\begin{equation}
    \phi_{bn} = \phi_{\text{ref}} + \frac{K\times \text{DM}}{P_s}\left(\frac{1}{\nu_n^{2}} - \frac{1}{\nu_\text{ref}^{2}}\right) - \delta_b\,,
    \label{eq:phase-shift-corr}
\end{equation}
where 
\begin{equation}
    \delta_b = \frac{t_{fb}-t_{f0}}{P_s}\,,  \label{eq:deltab}
\end{equation}
and we have arbitrarily chosen the band labeled $b=0$ as the reference and $P_s$ is the pulsar spin period\footnote{In this work, we are considering the period from the center of the observation}.
The frequency $\nu_\text{ref}$ is chosen such that the covariance between $\phi_{\text{ref}}$ and the  DM implied by equation \eqref{eq:wb-chisq-comb} vanishes. Note that this method \textit{preserves the full phase resolution} available in each band since the number of phase bins need not be equal for the different bands in equations \eqref{eq:chisq_funcs} and \eqref{eq:wb-chisq-comb}-\eqref{eq:deltab}.

\subsubsection{Wideband Timing with \tempotwo{} using \libstempo{}}  \label{timingprocedure}

The wideband likelihood was previously only available in \texttt{tempo} \citep{NiceDemorest+2015} and \texttt{PINT} \citep{LuoRansom+2021}. In this work, We implement the wideband likelihood using the \texttt{libstempo} \citep{Vallisneri2020} python interface to perform wideband timing in the \tempotwo{} framework (refer Appendix \ref{appB} for details).
We considered $\texttt{DMEFAC}$ and $\texttt{T2EFAC}$\footnote{\texttt{DMEFAC} and \texttt{T2EFAC} are white noise parameters used to scale the DM and ToA uncertainties,  respectively} to account for the radiometer noise contribution to the DM and ToA uncertainties, respectively. These are estimated via a $\chi^2$-implementation done with \libstempo{} and the optimum fit parameters for various pulsars were chosen as per the InPTA DR1 Narrowband timing \citep{TarafdarNobleson+2022}. The \texttt{DMEFAC} and \texttt{T2EFAC} values were estimated such that the reduced $\chi^2$ obtained by iteratively fitting the timing parameters is close to unity along with the post-fit weighted RMS to be of the order of a few 100s of ns to a few $\mu$s, for each of the Band 3, CC, and CP ToAs. In this way, ToAs obtained from the combination of data from two non-contiguous frequency bands are timed for the first time within the paradigm of the wideband technique.

\section{Application on PSR J1909--3744}
\label{sec:validation}
PSR J1909$-$3744 is a binary MSP with a rotational period $P_s\sim$2.95 ms. It was discovered using the Parkes 64-m Radio Telescope in the Swinburne High Latitude Pulsar Survey \citep{JacobyBailes+2003}. It is one of the best pulsars for PTA studies \citep{VerbiestLentati+2016,PereraDeCesar+2019} due to its sharp pulse profile, low-profile evolution with the radio frequency, and well-studied timing model \citep{LiuGuillemot+2020}.
Here, we demonstrate and validate the CC and CP methods (\S \ref{sec:wb-combine-methods}) using the uGMRT Band 3 and Band 5 data of PSR J1909$-$3744 from Cycles 37-40 (MJDs $58781-59496$), with 200 MHz bandwidth (BW), obtained as a part of the InPTA campaign \citep{TarafdarNobleson+2022}. We used MJD $59630$ as the template epoch obtained from InPTA observations of Cycle 41 of the uGMRT. The details of observations and data reduction procedures for these datasets can be found in \citet{SusobhananMaan+2021} and \citet{TarafdarNobleson+2022}.

\subsection{Combined Portrait (CP) method} \label{cpJ1909-3744}
As discussed in subsection \ref{subsec:cp_description}, the CP method requires the phase resolution of two bands to be the same for combining the data. The Band 5 uGMRT data of the InPTA campaign is configured to be recorded with a smaller time resolution than the Band 3 data, which leads to a smaller number of phase bins in Band 5 than in Band 3 when the data is folded. Hence, we phase-collapsed the Band 3 data to the same number of phase bins as those of Band 5 before appending the two bands using \texttt{psradd}.

\begin{figure}[ht]
   \centering
   \includegraphics[width=0.48\textwidth]{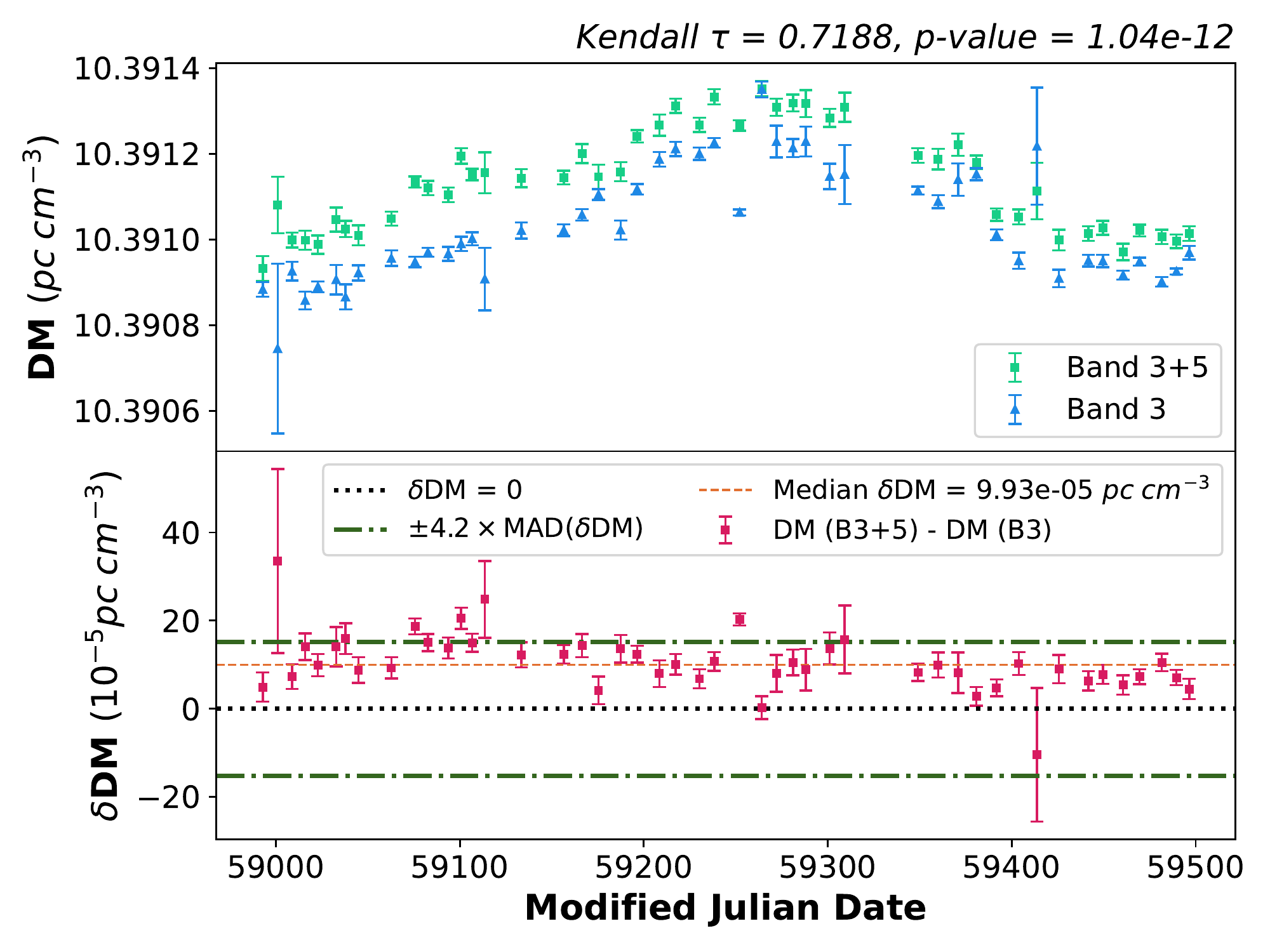}
   \caption{\small{Wideband DMs obtained from the Combined Portrait (CP) method for Band 3+5 data along with traditional Band 3 (single-band) results for PSR J1909$-$3744. The top panel shows the DM time series for  Band 3+5 CP (green points) and Band 3 (blue points) overlaid for comparison. The Kendall Tau coefficient on top right shows the correlation between the two time series with the mentioned $p$-value. The bottom panel shows the DM differences ($\delta$DM in units of $10^{-5}$\;\text{pc}\;\text{cm}$^{-3}$) on subtracting both time series (Band 3+5 $-$ Band 3) where the median value is shown by the dashed line and the dash-dotted lines representing the MAD-band (equivalent to $3\sigma$ contour).}}
   \label{fig:J1909-3744_m1_DM+DMresiduals}
\end{figure}

\begin{figure}[ht]
    \centering
    \includegraphics[width=0.48\textwidth]{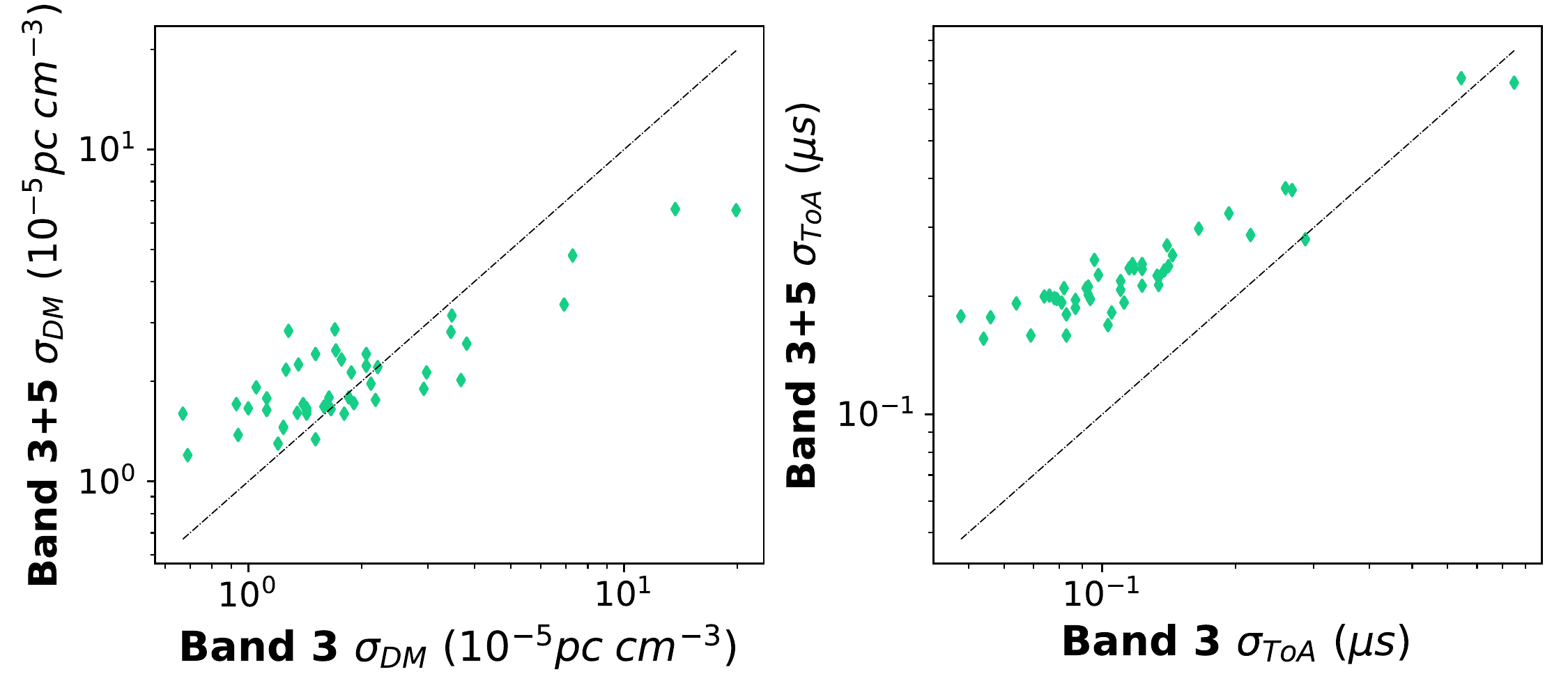}
   \caption{\small{A comparison of Wideband DM (left panel) and ToA (right panel) uncertainties obtained from the CP method for Band 3+5 data is shown against those obtained from the traditional Band 3 data for PSR J1909$-$3744.
   The Band 3+5 uncertainties estimated with the CP method are shown on the vertical axis and Band 3 uncertainties on the horizontal axis (in units of $10^{-5}\;\text{pc}\;\text{cm}^{-3}$ for DM uncertainties, and in units of $\mu\text{s}$ for ToA uncertainties). The diagonal dash-dotted line shows the $y=x$ curve. Points lying below the diagonal line show that the obtained uncertainties with the CP method are lower compared to Band 3 results and vice versa. }}
   \label{fig:J1909-3744_m1_DME+ToAE}
\end{figure}

A comparison of wideband DM time series of Band 3+5 (CP) and Band 3 is shown in figure \ref{fig:J1909-3744_m1_DM+DMresiduals}, where the Band 3 DM time series is obtained while preserving the original phase resolution. The Kendall Tau correlation coefficient \citep{Kendall1938} of value 0.7188 and $p$-value $\sim\times 10^{-12}$ indicates a good agreement between the two DM time series. We also see a slight offset between Band 3 and Band 3+5 CP method DM time series (refer section \ref{DMtimeseries} for a discussion). In figure \ref{fig:J1909-3744_m1_DME+ToAE}, we compare the DM (left panel) and ToA (right panel) precisions of Band 3+5 (CP) and Band 3 time series. The points lying below the dashed diagonal line indicate an improved DM or ToA precision with the CP method compared to Band 3 results and vice versa. 

We see in figure \ref{fig:J1909-3744_m1_DME+ToAE} that all epochs do not show an improved DM precision, and most of the epochs show a worsened ToA precision i.e., we found a decrement in the median precision or an increase in the median uncertainties values ($\sigma_{\text{DM}}$ and $\sigma_{\text{ToA}}$) of Band 3+5 (CP) results compared to Band 3. This is primarily due to the decreased phase resolution of Band 3 data used in the CP method. Hence, for combining bands, the CP method has a disadvantage, especially for MSPs like J1909$-$3744, wherein the pulse profile is sharp with minimal features, leading to only a few phase bins in the pulse region of the profile upon toning down the phase resolution which leads to a loss of information content,  and thereby leading to poor template construction as well as bad DM and ToA estimates. 

\subsection{Combined Chi-squared (CC) method} \label{ccJ1909-3744}
\begin{figure}[ht]
    \centering
    \includegraphics[width=0.48\textwidth]{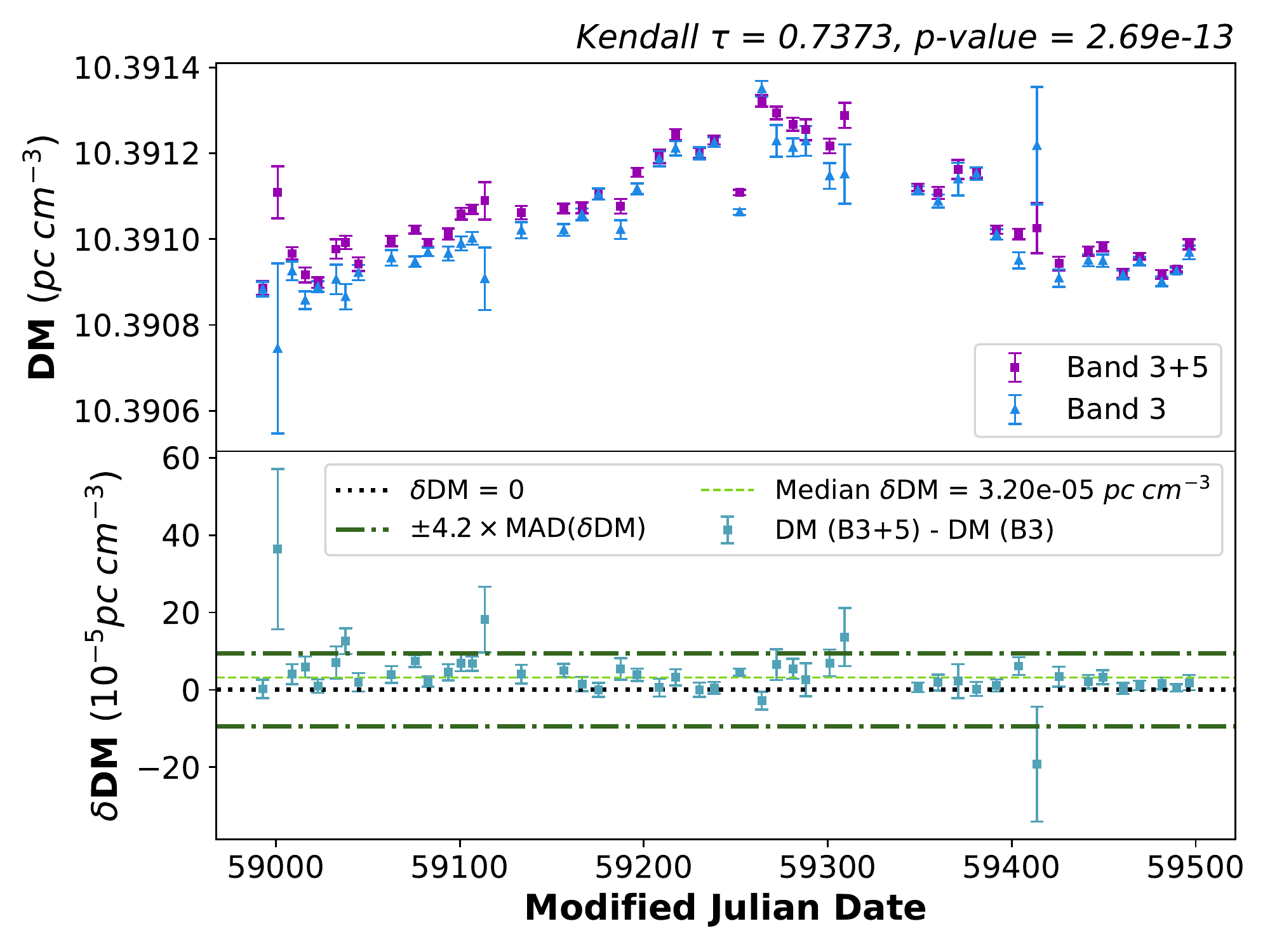}
    \caption{\small{Wideband DMs obtained from the Combined Chi-squared (CC) method for Band 3+5 data along with traditional Band 3 (single-band) results for PSR J1909$-$3744. The top panel shows the DM time series for  Band 3+5 CC (magenta points) and Band 3 (blue points) overlaid for comparison. The Kendall Tau coefficient on top right shows the correlation between the two time series with the mentioned p-value. The bottom panel shows the DM differences ($\delta$DM in units of $10^{-5}\;\text{pc}\;\text{cm}^{-3}$) on subtracting both time series (Band 3+5 $-$ Band 3), where the median value is shown by the dashed line and the dash-dotted lines representing the MAD-band (equivalent to $3\sigma$ contour).}}
    \label{fig:J1909-3744_m2_DM+DMresiduals}
\end{figure}

\begin{figure}[ht]
    \centering
    \includegraphics[width=0.48\textwidth]{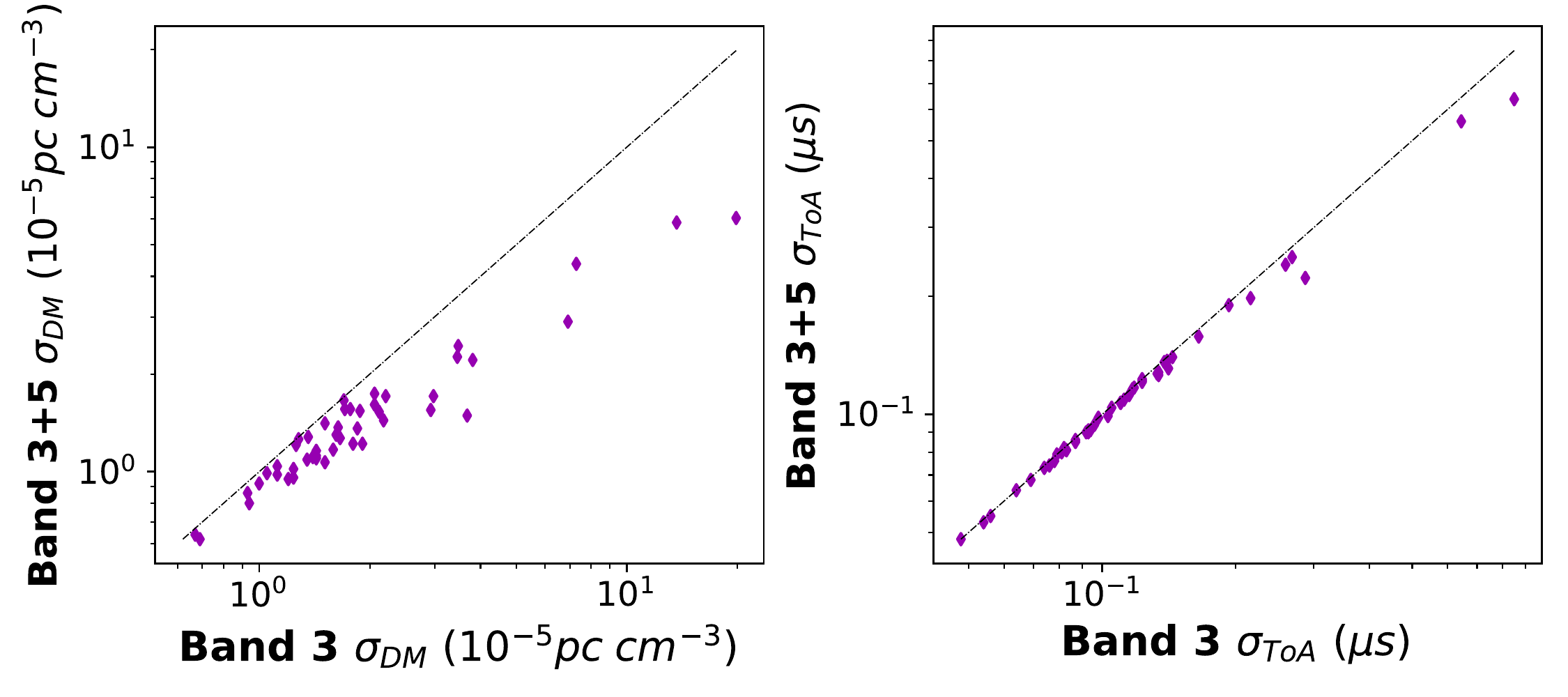}
    \caption{\small{A comparison of Wideband DM (left panel) and ToA (right panel) uncertainties obtained from the CC method for Band 3+5 data is shown against those obtained from the traditional Band 3 data for PSR J1909$-$3744. The Band 3+5 uncertainties estimated with the CC method are shown on the vertical axis and Band 3 uncertainties on the horizontal axis (in units of $10^{-5}\;\text{pc}\;\text{cm}^{-3}$ for DM uncertainties, and in units of $\mu\text{s}$ for ToA uncertainties). The diagonal dash-dotted line shows the $y=x$ curve. Points lying below the diagonal line show that the obtained uncertainties with the CC method are lower compared to Band 3 results and vice versa.}}
    \label{fig:J1909-3744_m2_DME+ToAE}
\end{figure}

The CC method preserves the native phase resolution of Band 3 and Band 5 data as well as the template portraits, since it incorporates them within a combined Fourier domain $\chi^2$-statistic as described in subsection \ref{subsec:cc_description}. Figure \ref{fig:J1909-3744_m2_DM+DMresiduals} shows the Band 3+5 DM time series obtained using the CC method in comparison with the Band 3 DM time series, wherein we can see that the Band 3+5 DMs bear a high positive correlation with the Band 3 DMs, showing a good agreement between the two. Figure \ref{fig:J1909-3744_m2_DME+ToAE} shows the Band 3+5 (CC) DM (left panel) and ToA (right panel) uncertainties in comparison with the Band 3 results. We can see that the Band 3+5 $\sigma_{\text{DM}}$ values are smaller than those of Band 3 (all lying below the $y=x$ curve), hence showing a universal improvement in the median DM precision after band-combination. The Band 3+5 $\sigma_{\text{ToA}}$ values are also slightly less compared to Band 3 leading to an improvement in the median ToA precision as well.

Overall, the CC method provides significant improvements for Band 3+5 compared to Band 3 results, especially because of preserving the native Band 3 phase resolution, unlike the CP method. The templates are also more effectively modelled because of applying PCA separately on each band without the need for interpolating over a large frequency gap of ~$760$ MHz between Band 3 and Band 5. Similarly, when we apply PCA to model the template for the CP method on the Band 3+5 data obtained using \texttt{psradd}, there is a possibility that the PCA method may not be interpolating the profile evolution accurately due to the wide band separation ($\sim 760$ MHz) between Band 3 and Band 5 data. Therefore, CC method comes out as a more robust method for the  band combination.

\subsection{Split-band test for the CC method} \label{splitJ1909-3744}
\begin{figure}[ht]
    \centering
    \includegraphics[width=0.48\textwidth]{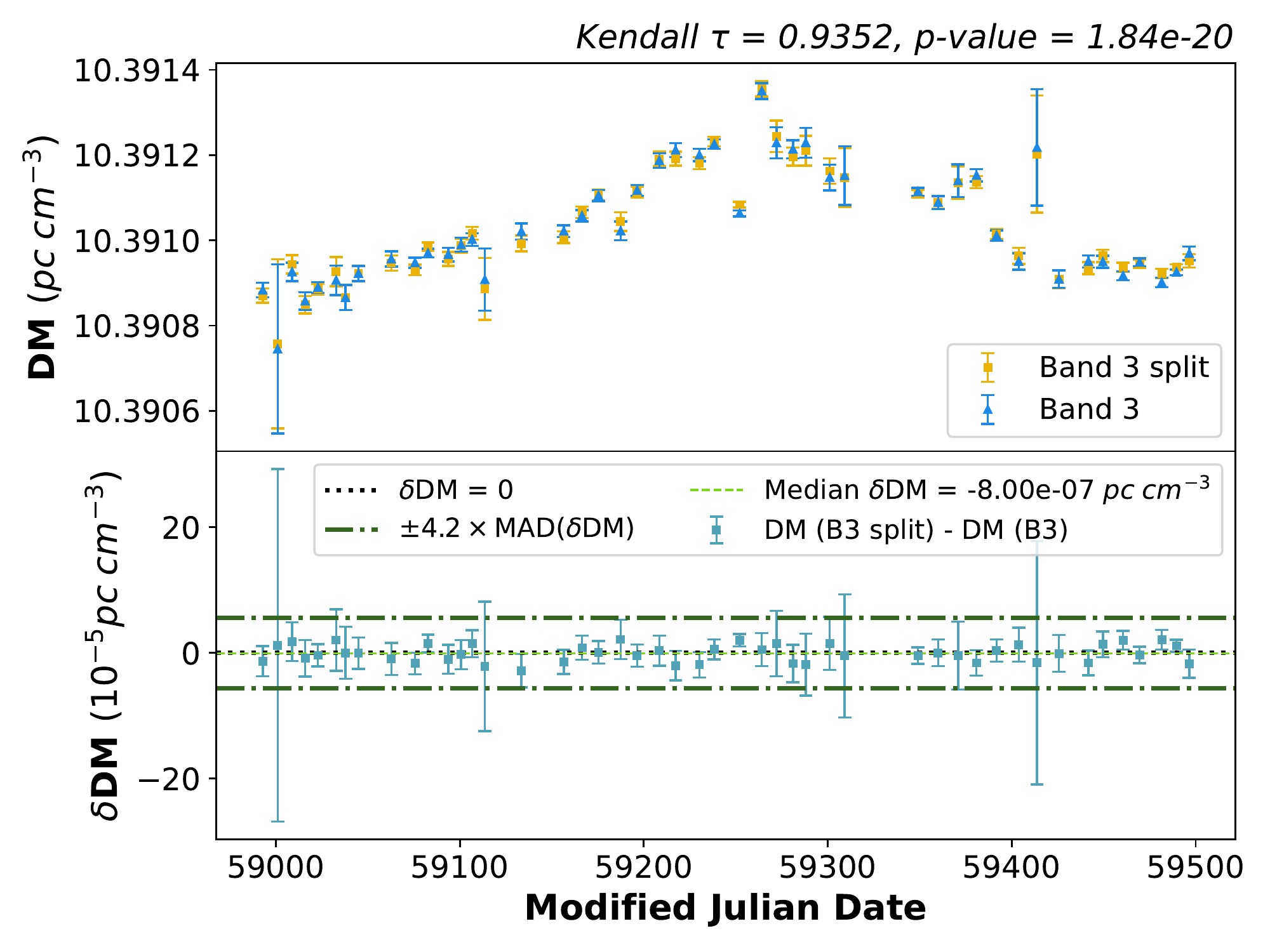}
    \caption{\small{The split-band test results for PSR J1909$-$3744. The top panel shows the wideband DM time series for Band 3 alone (blue points), and by splitting it into two 100 MHz bands and using the CC method on the two sub-bands (yellow points) overlaid for comparison. The Kendall Tau coefficient on top right shows the correlation between the two time series with the mentioned p-value. The bottom panel shows the DM differences ($\delta$DM in units of $10^{-5}\;\text{pc}\;\text{cm}^{-3}$) on subtracting both time series (Band 3 split $-$ Band 3 single-band) with the median value shown by the dashed line and the dash-dotted lines representing the MAD-band (equivalent to $3\sigma$ contour).}}
    \label{fig:J1909-3744_split_band_test}
\end{figure}

To validate the application of our novel CC method to combine the data of two bands for estimating wideband DMs and ToAs, we perform a split-band test. In this test, we consider one of the 200 MHz BW data (Band 3 is selected as it has higher S/N than Band 5) and split it into two sub-bands each with a bandwidth of 100 MHz using the \texttt{psrsplit} command of \texttt{psrchive}. We then estimate the DM time series obtained by applying the CC method on these two sub-bands and compare it with the wideband DM estimates obtained for the full 200 MHz BW data. The split-band test results for PSR J1909-3744 are shown in figure \ref{fig:J1909-3744_split_band_test}, where we can see that the DM values are in close agreement with Kendall Tau value $\tau \sim 0.94$ and $p\sim 2\times 10^{-20}$ implying strong (positive) correlation with the single band result for Band 3. The strong (positive) correlation with negligible offsets indicates that the CC method for combining bands, within the regime of the wideband technique, is working well. Hence, the split-band test serves as a litmus test for validating the new technique.
 
\subsection{Wideband Timing results for PSR J1909\texorpdfstring{$-$}{-}3744}
\label{timingJ1909-3744}
\begin{figure}[ht]
    \centering
    \includegraphics[width=\linewidth]{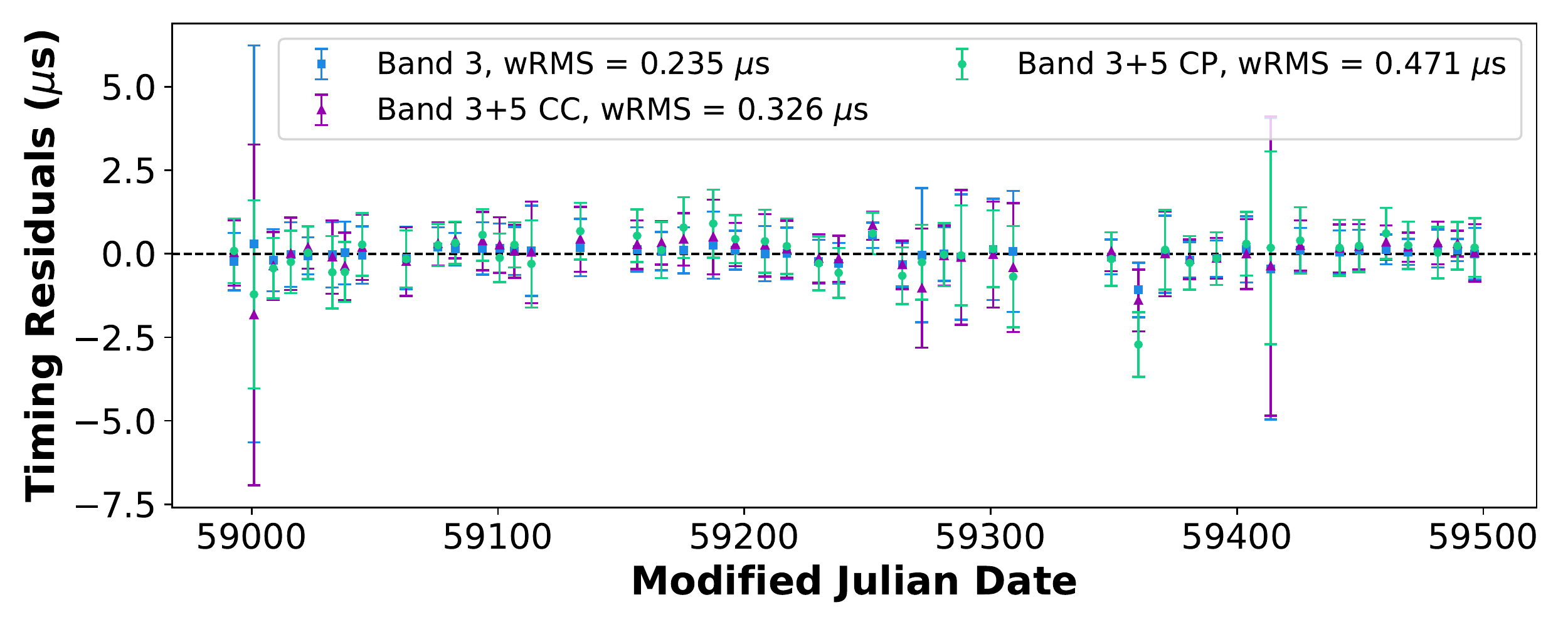}
    \caption{\small{Timing results for PSR J1909-3744 obtained by implementing the wideband likelihood inside \texttt{tempo2} using \texttt{libstempo}. The \texttt{DMEFAC} and \texttt{T2EFAC} values are estimated separately for the cases of Band 3 (blue points), CC (magenta points) and CP (green points) ToAs with only $200$ MHz data. The error bars shown in the plot have \texttt{T2EFAC} values incorporated. The post-fit weighted RMS ToA residual values for the Band 3, CC, and CP timing residuals are shown in the top right corner and the dashed horizontal line corresponds to zero residual level.}}
    \label{fig:J1909-3744_timing_results}
\end{figure}

The wideband timing results for PSR J1909$-$3744 obtained by implementing the wideband likelihood with \texttt{tempo2} using \texttt{libstempo} are shown in figure \ref{fig:J1909-3744_timing_results}. The \texttt{DMEFAC} and \texttt{T2EFAC} values are estimated for each of the Band 3, CC, and CP ToAs. We then incorporate the \texttt{T2EFAC} and \texttt{DMEFAC} values to generate a global timing solution. The post-fit timing residuals obtained from this procedure are shown in figure \ref{fig:J1909-3744_timing_results}. The post-fit weighted RMS ToA residual values for the Band 3, CC, and CP timing residuals are obtained to be 0.235 $\mu$s, 0.326 $\mu$s, and 0.471 $\mu$s respectively, and are consistent with each other. We fit the same parameters as fitted in the InPTA DR1 \citep{TarafdarNobleson+2022} narrowband timing method, which are \texttt{F0} and \texttt{F1} for the case of J1909$-$3744. 

\begin{table*}
\centering
\begin{tabular}{c|ccc}
\hline
\hline
\textbf{\begin{tabular}[c]{@{}c@{}}Parameter\\Name (unit)\end{tabular}} &
\textbf{\begin{tabular}[c]{@{}c@{}}Band 3\\ {\begin{tabular}{cc} Value & Uncertainty\end{tabular}} \end{tabular}} &
\textbf{\begin{tabular}[c]{@{}c@{}}Band 3+5 CC\\ {\begin{tabular}{cc} Value & Uncertainty\end{tabular}} \end{tabular}} &
\textbf{\begin{tabular}[c]{@{}c@{}}Band 3+5 CP\\ {\begin{tabular}{cc} Value & Uncertainty\end{tabular}} \end{tabular}} \\ \hline \\
F0 ($\text{s}^{-1}$) \space\space & {\begin{tabular}{cc} $339.31568666042$ \space\space & $3.1\times10^{-11}$ \end{tabular}} & {\begin{tabular}{cc} $339.31568666042$ \space\space & $2.6\times10^{-11}$ \end{tabular}} & {\begin{tabular}{cc} $339.31568666042$ \space\space & $1.8\times10^{-11}$ \end{tabular}} \\ \\
F1 ($\text{s}^{-2}$) \space\space & {\begin{tabular}{cc} $-1.615\times10^{-15}$ \space\space & $1.3\times10^{-18}$ \end{tabular}} & {\begin{tabular}{cc} $-1.615\times10^{-15}$ \space\space & $1.1\times10^{-18}$ \end{tabular}} & {\begin{tabular}{cc} $-1.6158\times10^{-15}$ \space\space & $8.1\times10^{-19}$ \end{tabular}} \\ \\ \hline\hline
\end{tabular}
\caption{Table of fitted parameters obtained from the wideband timing of PSR J1909$-$3744 using \texttt{tempo2}. The first column specifies the fitted parameters. The second, third, and fourth columns list the Band 3, Band 3+5 CC, and Band 3+5 CP timing results respectively, enlisting the parameter values and their uncertainties. }
\label{tab:1}
\end{table*}

The post-fit RMS values for Band 3 and Band 3+5 (both CC and CP) are obtained to be very close, while the fitted parameters estimated using the Band 3+5 methods are having better precision compared to the former, as shown in table \ref{tab:1}. This clearly highlights the improvement in timing precision attained with Band 3+5 wideband timing, especially the CC method.

\section{Application on InPTA Data}
\label{sec:inpta-data-appl}
We now present the results obtained by implementing the CP and CC methods, and the wideband timing technique using \tempotwo{} on the InPTA dual-band data (Band 3: 300$-$500 MHz and Band 5: 1260$-$1460 MHz) of 14 MSPs. The same data was used for the first data release of InPTA (InPTA DR1: \cite{TarafdarNobleson+2022}). 

\begin{table*}
\centering
\begin{tabular}{c|ccc}
\hline \\
[-0.1cm]
& & \textbf{DM uncertainties ($\text{pc}\;\text{cm}^{-3}$)} \\
[0.2cm]
\hline
\textbf{\begin{tabular}[c]{@{}c@{}}Pulsar\\Name\end{tabular}} &
\textbf{\begin{tabular}[c]{@{}c@{}}Band 3\\{\begin{tabular}{cc} Median & Minimum \end{tabular}} \end{tabular}} &
\textbf{\begin{tabular}[c]{@{}c@{}}Band 3+5 CC\\{\begin{tabular}{cc} Median & Minimum \end{tabular}} \end{tabular}} &
\textbf{\begin{tabular}[c]{@{}c@{}}Band 3+5 CP\\ {\begin{tabular}{cc} Median & Minimum \end{tabular}} \end{tabular}} \\ \hline \\ 
J0437$-$4715 & {\begin{tabular}{cc} $1.8\times10^{-4}$ & $1.2\times10^{-4}$ \end{tabular}} & {\begin{tabular}{cc} $4.2\times10^{-5}$ & $3.7\times10^{-5}$ \end{tabular}} & {\begin{tabular}{cc} $4.1\times10^{-5}$ & $3.7\times10^{-5}$ \end{tabular}} \\ \\
J0613$-$0200 & {\begin{tabular}{cc} $7.8\times10^{-5}$ & $3.2\times10^{-5}$ \end{tabular}} & {\begin{tabular}{cc} $5.6\times10^{-5}$ & $2.9\times10^{-5}$ \end{tabular}} & {\begin{tabular}{cc} $6.0\times10^{-5}$ & $3.1\times10^{-5}$ \end{tabular}} \\ \\
J0751$+$1807 & {\begin{tabular}{cc} $4.1\times10^{-4}$ & $1.9\times10^{-4}$ \end{tabular}} & {\begin{tabular}{cc} $2.0\times10^{-4}$ & $1.0\times10^{-4}$ \end{tabular}} & {\begin{tabular}{cc} $2.1\times10^{-4}$ & $1.0\times10^{-4}$ \end{tabular}} \\ \\
J1012$+$5307 & {\begin{tabular}{cc} $5.9\times10^{-5}$ & $1.9\times10^{-5}$ \end{tabular}} & {\begin{tabular}{cc} $3.8\times10^{-5}$ & $1.8\times10^{-5}$ \end{tabular}} & {\begin{tabular}{cc} $4.0\times10^{-5}$ & $2.4\times10^{-5}$ \end{tabular}} \\ \\
J1022$+$1001 & {\begin{tabular}{cc} $1.1\times10^{-4}$ & $4.8\times10^{-5}$ \end{tabular}} & {\begin{tabular}{cc} $9.8\times10^{-5}$ & $4.8\times10^{-5}$ \end{tabular}} & {\begin{tabular}{cc} $1.0\times10^{-4}$ & $4.8\times10^{-5}$ \end{tabular}} \\ \\
J1600$-$3053 & {\begin{tabular}{cc} $2.1\times10^{-4}$ & $1.3\times10^{-4}$ \end{tabular}} & {\begin{tabular}{cc} $7.8\times10^{-5}$ & $6.1\times10^{-5}$ \end{tabular}} & {\begin{tabular}{cc} $8.4\times10^{-5}$ & $5.7\times10^{-5}$ \end{tabular}} \\ \\
J1643$-$1224 & {\begin{tabular}{cc} $1.2\times10^{-4}$ & $6.4\times10^{-5}$ \end{tabular}} & {\begin{tabular}{cc} $6.2\times10^{-5}$ & $3.6\times10^{-5}$ \end{tabular}} & {\begin{tabular}{cc} $6.3\times10^{-5}$ & $3.4\times10^{-5}$ \end{tabular}} \\ \\
J1713$+$0747 & {\begin{tabular}{cc} $7.3\times10^{-5}$ & $2.8\times10^{-5}$ \end{tabular}} & {\begin{tabular}{cc} $3.2\times10^{-5}$ & $1.8\times10^{-5}$ \end{tabular}} & {\begin{tabular}{cc} $4.1\times10^{-5}$ & $2.2\times10^{-5}$ \end{tabular}} \\ \\
J1744$-$1134 & {\begin{tabular}{cc} $2.6\times10^{-5}$ & $1.5\times10^{-5}$ \end{tabular}} & {\begin{tabular}{cc} $1.9\times10^{-5}$ & $8.7\times10^{-6}$ \end{tabular}} & {\begin{tabular}{cc} $2.7\times10^{-5}$ & $1.2\times10^{-5}$ \end{tabular}} \\ \\
J1857$+$0943 & {\begin{tabular}{cc} $2.0\times10^{-4}$ & $7.2\times10^{-5}$ \end{tabular}} & {\begin{tabular}{cc} $8.7\times10^{-5}$ & $3.5\times10^{-5}$ \end{tabular}} & {\begin{tabular}{cc} $9.1\times10^{-5}$ & $3.6\times10^{-5}$ \end{tabular}} \\ \\
J1909$-$3744 & {\begin{tabular}{cc} $1.6\times10^{-5}$ & $6.7\times10^{-6}$ \end{tabular}} & {\begin{tabular}{cc} $1.3\times10^{-5}$ & $6.2\times10^{-6}$ \end{tabular}} & {\begin{tabular}{cc} $1.8\times10^{-5}$ & $1.2\times10^{-5}$ \end{tabular}} \\ \\
J1939$+$2134 & {\begin{tabular}{cc} $2.8\times10^{-6}$ & $1.1\times10^{-6}$ \end{tabular}} & {\begin{tabular}{cc} $2.7\times10^{-6}$ & $1.1\times10^{-6}$ \end{tabular}} & {\begin{tabular}{cc} $1.7\times10^{-5}$ & $2.6\times10^{-6}$ \end{tabular}} \\ \\
J2124$-$3358 & {\begin{tabular}{cc} $1.3\times10^{-4}$ & $2.0\times10^{-5}$ \end{tabular}} & {\begin{tabular}{cc} $1.1\times10^{-4}$ & $2.0\times10^{-5}$ \end{tabular}} & {\begin{tabular}{cc} $1.3\times10^{-4}$ & $2.2\times10^{-5}$ \end{tabular}} \\ \\
J2145$-$0750 & {\begin{tabular}{cc} $3.3\times10^{-5}$ & $1.0\times10^{-5}$ \end{tabular}} & {\begin{tabular}{cc} $2.5\times10^{-5}$ & $1.0\times10^{-5}$ \end{tabular}} & {\begin{tabular}{cc} $2.5\times10^{-5}$ & $1.0\times10^{-5}$ \end{tabular}} \\ \\ \hline\hline
\end{tabular}
\caption{Table of DM uncertainties (in units of $\text{pc}\;\text{cm}^{-3}$). The first column specifies the pulsar names. The second column lists the median and minimum errors in the DM estimation using the standard Wideband technique on Band 3 (single-band). The third and fourth columns enlist the median and minimum DM errors using the wideband CC and CP methods respectively. All listed values are calculated by including both 100 MHz and 200 MHz bandwidth data.}
\label{tab:2}
\end{table*}

\subsection{Description of InPTA DR1} \label{dr1_desc}

The InPTA DR1 \citep{TarafdarNobleson+2022} constitutes 3.5 years of data corresponding to the observations of 14 MSPs obtained using the uGMRT \cite[][]{GuptaAjithkumar+2017}. The data spans from 2018 to 2021 and has a typical cadence of two weeks, carried out during uGMRT observing cycles 34$-$35 and 37$-$40. These observations were performed by dividing the 30 uGMRT antennae into multiple phased subarrays which were used to observe the same source in multiple frequency bands simultaneously. The data were recorded in total intensity mode \citep{JoshiGopakumar+2022}.
The GMRT Wideband Backend \cite[GWB:][]{ReddyKudale+2017} was used to record the channelized time series data in binary format, and then RFI-mitigated and reduced to \psrfits{} archives using the \pinta{} pipeline \cite[][]{SusobhananMaan+2021}. During cycles 34$-$35 we observed MSPs simultaneously in Band 3 (400$-$500 MHz), Band 4 (650$-$750 MHz) and Band 5 (1360$-$1460 MHz) of uGMRT with 100 MHz bandwidth in each band. During cycles 37$-$40, we performed simultaneous observations only in Band 3 (300$-$500 MHz) and Band 5 (1260$-$1460 MHz) with 200 MHz bandwidth. The Band 3 data in all cycles as well as the Band 5 data in cycles 34$-$35 (except observations between Oct. 20, 2018 and
Nov. 14, 2018) were coherently dedispersed using a real-time
pipeline \citep{dg2016} to the known DM of each pulsar. uGMRT can perform coherent dedispersion on a total bandwidth of 200 MHz only, so in cycles 34$-$35, observations were made with 100 MHz bandwidth in each band so that both Band 3 and Band 5 data can be coherently dedispersed \citep{TarafdarNobleson+2022}.

The Global Positioning System (GPS) was used to measure the narrowband ToAs and the hydrogen maser clock at the uGMRT provided a local topocentric frequency standard.
The narrowband timing residuals in the InPTA DR1 were obtained using \tempotwo{} \cite[][]{HobbsEdwardsManchester2006}. 
The timing residuals were also generated from the wideband likelihood method \cite[][]{PennucciDemorestRansom2014, Pennucci2019, AlamArzoumanian+2021, NoblesonAgarwal+2022} using \texttt{TEMPO} \cite[][]{NiceDemorest+2015} for Band 3 data only. 
The epoch-wise DM corrections were introduced in the fit. 
The DMX parameters were calculated from the DM time series estimated using \dmcalc{} \citep{KrishnakumarManoharan+2021} for the narrowband timing from low-frequency uGMRT data obtained in Band 3 and Band 5 simultaneously. Similarly, DMX parameters were estimated using the wideband likelihood method for wideband timing from Band 3 data of the uGMRT.

\subsection{DM time series} \label{DMtimeseries}
 
\begin{figure*}
    \centering
    \includegraphics[scale=1.1, width=\linewidth]{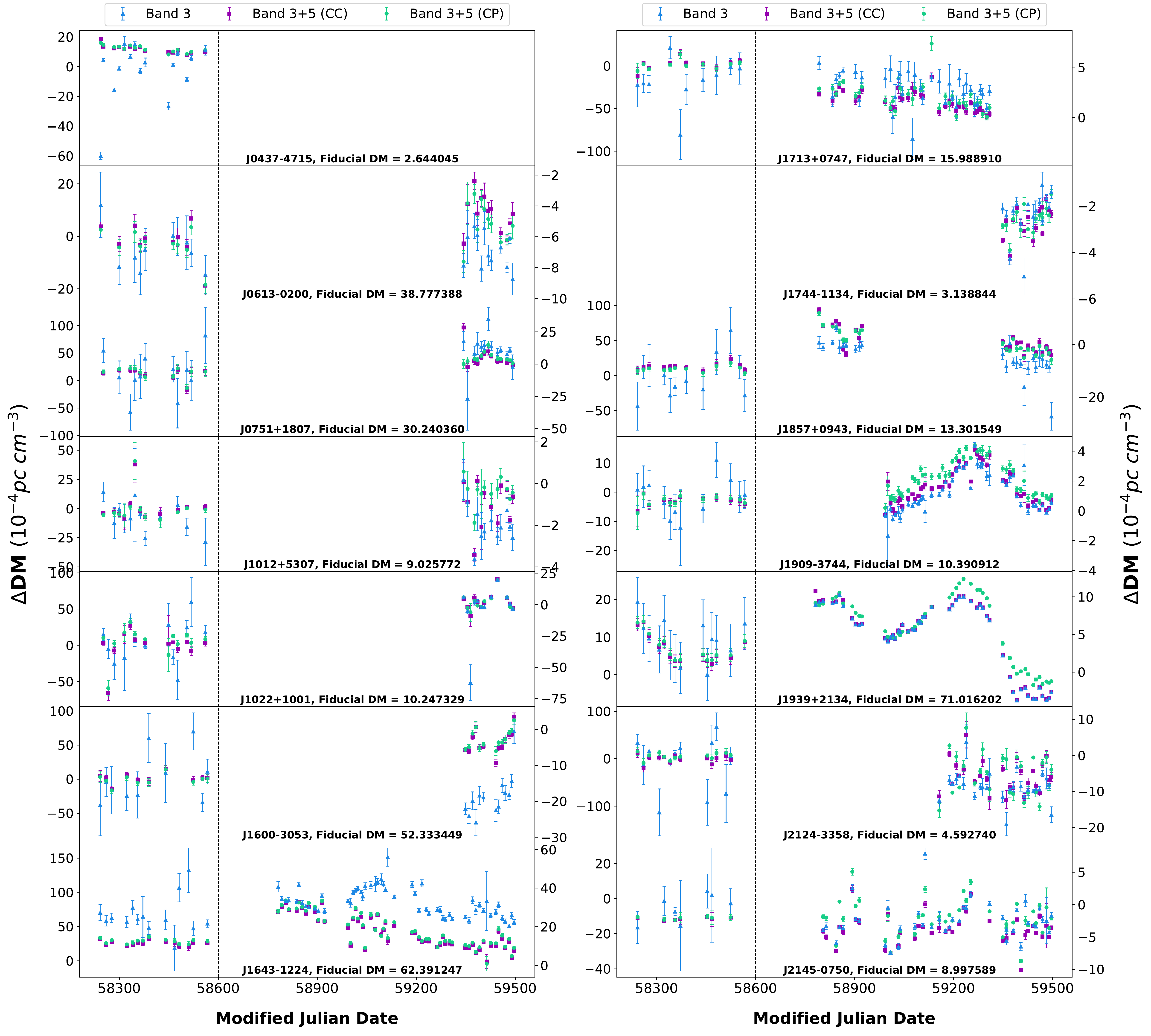}
    \caption{Consolidated DM time series for 14 InPTA DR1 pulsars. The vertical axes depict the difference ($\Delta\text{DM}$ in units of $10^{-4}\;\text{pc}\;\text{cm}^{-3}$) between the fiducial DM and the corresponding estimated DMs for each pulsar estimated by applying the standard wideband technique on Band 3 data (blue points), using the wideband CC method (magenta points) and the wideband CP method (green points) on Band 3+5 data. The horizontal axes depict epochs in terms of the Modified Julian Date. The estimated DM precision of 200 MHz bandwidth data is higher than that of 100 MHz bandwidth data, hence the horizontal axes are split into two parts at MJD 58600 with dashed vertical lines, with 100 MHz bandwidth epochs on the left side and 200 MHz bandwidth epochs on the right side of the dashed line. The vertical axes in each panel are scaled differently for 100 MHz (left axis) and 200 MHz (right axis) epochs such that the DM variations are clearly visible. Pulsar names and their respective fiducial DM values are mentioned at the bottom of each respective panel.}
    \label{fig:dm_comp_plot}
\end{figure*}

In this section, we present the wideband DM time series obtained for 14 InPTA DR1 pulsars using the CC and CP methods, described in sections \ref{subsec:cp_description} and \ref{subsec:cc_description}, for the combination of Band 3 and Band 5 InPTA data. We compare our Band 3+5 combination results with the Band 3 (single-band) DM time series of these pulsars estimated using the standard wideband method (described in section \ref{sec:wb-overview}). The template epochs used for ToA and DM estimation are the same as those used in the InPTA DR1 analysis for all pulsars, which are high-S/N epochs selected from Cycle 41 of the uGMRT for the respective pulsars. We also keep the template epoch to be the same for Band 3 alone, the CC, and the CP analysis to maintain consistency. In table \ref{tab:2}, we have listed the median and minimum uncertainties in DMs estimated for (i) Band 3, (ii) Band 3+5 CC, and (iii) Band 3+5 CP, including both 100 MHz and 200 MHz data. It is evident from the listed uncertainty values that there is a significant improvement in the DM precision when Band 3 and Band 5 data is combined using the CC method. However, for the CP method we find that the median DM precision goes slightly down compared to Band 3 for PSRs J1744$-$1134, J1909$-$3744, and J1939$+$2134, while remains same for PSRs J2124$-$3358. Overall, there is nearly two times increment in median DM precision using CC method for most of the pulsars.

A consolidated DM time-series plot illustrating the epoch-by-epoch DM
variations for all 14 InPTA DR1 pulsars is presented in figure \ref{fig:dm_comp_plot}. The plot shows the Band 3, CC, and CP method results, where the vertical axes in both panels depict the differences between the estimated DMs and the fiducial DMs. The DM precisions estimated from  100 MHz bandwidth (BW) data are lower as compared to those obtained from the  200 MHz bandwidth (BW) data, hence the scaling of the vertical axes is made separately for these two cases to make the DM variations over both 100 MHz and 200 MHz bandwidth epochs clearly visible. The epochs having these two different bandwidths are separated along the horizontal axis with a vertical dashed line at MJD 58600. The fiducial DM value for each pulsar is mentioned inside the respective panel of the figure. We have taken fiducial DM from the InPTA DR1 analysis. Refer \citep{TarafdarNobleson+2022} for more details.

 In both CC and CP methods, we are combining over a large gap in frequency that can cause differences in template portrait computations which are reflected as systematic DC offsets in DM time-series of Band3, CC, and CP methods, as seen in figure \ref{fig:dm_comp_plot}. A similar DM offset was also seen in the InPTA DR1 DM time-series estimated from the narrowband and wideband analysis, which was found to be caused by different templates used in the techniques~\citep{TarafdarNobleson+2022}. Here, in the CP method, we first \texttt{psradd} Band 3 and Band 5 data and then create an analytic template using the standard wideband technique, which means that the spline interpolation is done over a band-gap of $\sim$760 MHz. Whereas in the CC method, we supply separate analytic wideband templates of Band 3 and Band 5 which are internally used within the combined chi-square metric to estimate DMs and ToAs. This leads to the selection of different number of eigenprofiles and tolerance values \citep{PennucciDemorestRansom2014,Pennucci2019} in the CC and CP methods, leading to different analytical templates.  

We have provided a series of plots for 14 InPTA DR1 pulsars in appendix \ref{appA} to show a comparison between DM uncertainties estimated for Band 3 alone and Band 3+5 data with CC and CP methods. As the data of 100 and 200 MHz bandwidth have different sensitivities, they have different scales of corresponding uncertainties, hence we have presented them in different panels for each pulsar. In the case of 100 MHz bandwidth data, we see a significant improvement in the median DM precision for all pulsars with Band 3+5 data using both CC and CP methods compared to Band 3 alone results. For the 200 MHz bandwidth data, the CC method shows much higher improvement in the median DM precision than the CP method for all pulsars except J0437$-$4715. For PSRs J1744$-$1134, J1909$-$3744, and J1939$+$2134, we find a decrement in the median DM precision using the CP method compared to Band 3 alone for 200 MHz data, whereas the CC method shows improvement for these pulsars also. Such decrement in DM precision using CP method is expected due to reduced phase resolution in Band 3 which affects pulsars with sharp pulse profiles as explained in subsection \ref{cpJ1909-3744}. There is also a frequency gap of $\sim 750$ MHz between Band 3 and Band 5 data which affects the modeling of profile evolution across band edges in CP method, hence altering the results of pulsars with high profile evolution with radio frequency. For PSR J1643$-$1224, we observe that the trend in DM timeseries is not in agreement betweeen Band 3 and CC or CP methods. This effect can be explained in terms of scattering variations. PSR J1643-1224 has a highly scattered profile, especially at low radio frequencies. At widely separated radio frequencies, scattered pulses sample different path lengths through the ISM, which manifests as distinct variations in DMs \citep{2018MNRAS.479.4216M,Jaikombha,2016ApJ...817...16C,2019ApJ...878..130K} estimated for Band 3 and combination of Band 3+5 using CC or CP methods as seen in figure \ref{fig:dm_comp_plot}.

\subsection{ToAs and Timing residuals}
We show a comparison of ToA uncertainties estimated for Band 3 alone and Band 3+5 data with CC and CP methods in a series of plots for 14 InPTA DR1 pulsars in appendix \ref{appA}.
Similar to DM precision, we see a significant improvement in the median ToA precision for 100 MHz bandwidth data of all pulsars with Band 3+5 data using both CC and CP methods compared to Band 3 alone data. In the case of 200 MHz bandwidth data, there is improvement in median ToA precision using the CC method for PSRs J0751$+$1807, J1012$+$5307, J1600$-$3053, J1643$-$1224, J1713$+$0747, J1744$-$1134, J1857$+$0943 and J2145$-$0750 while it stays at par with Band 3 results for other pulsars. As the ToA precision depends on the S/N, and as Band 5 S/N is comparatively lesser than Band 3 S/N, therefore ToAs obtained after band combination, i.e. CC or CP ToAs, are not able to achieve a significant improvement in ToA uncertainty for 200 MHz data of all pulsars. The CP method shows improvement in median ToA precision than Band 3 alone for PSRs J0751$+$1807, J1012$+$5307, J1022$+$1001, J1600$-$3053, J1643$-$1224, and J2145$-$0750, whereas it decreases ToA precision for all other pulsars which can be attributed to the aforementioned reasons.

\begin{figure*}
    \centering
    \includegraphics[scale=1.1, width=\linewidth]{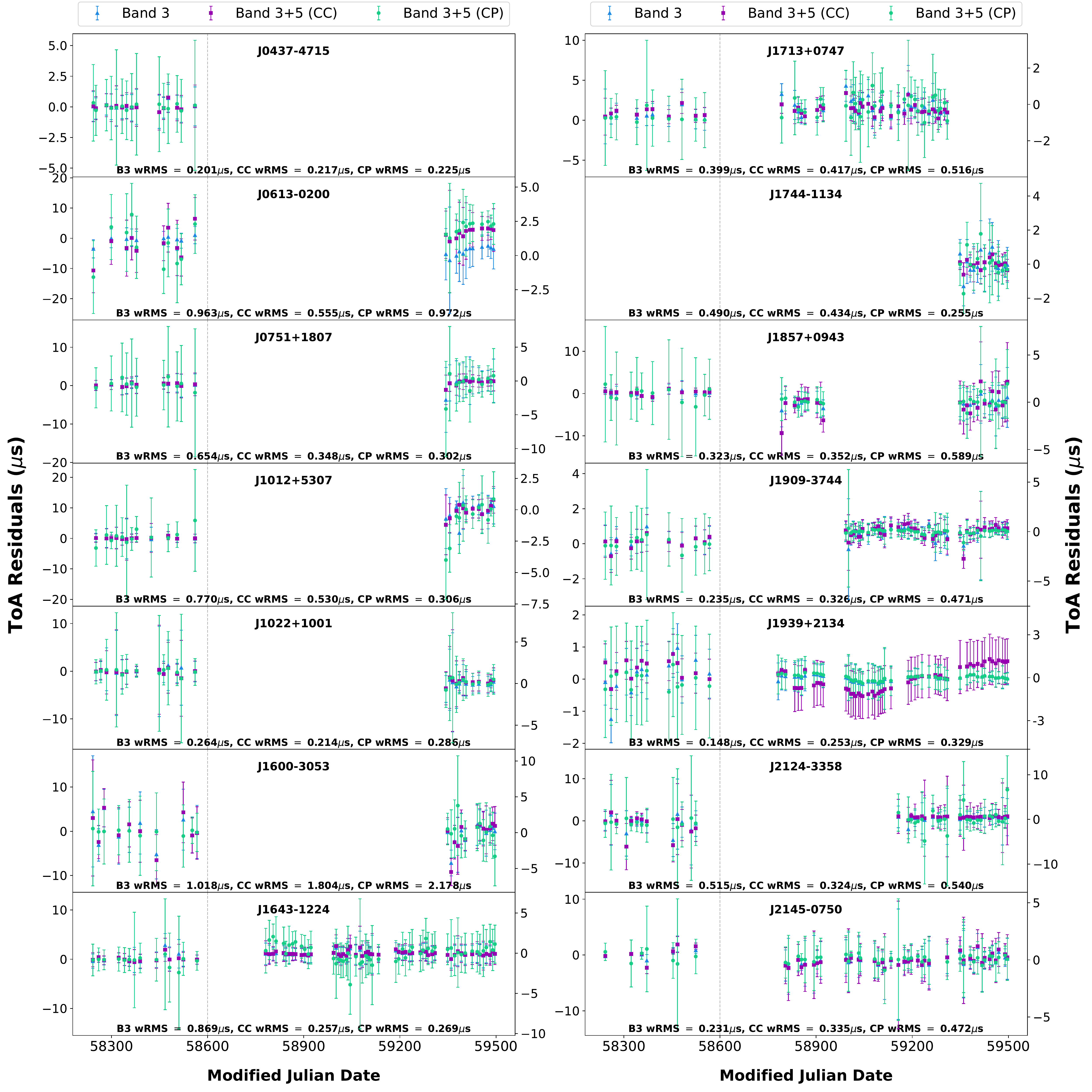}
    \caption{Consolidated wideband timing residuals for 14 InPTA DR1 pulsars. The post-fit wideband timing residuals (in units of $\mu\text{s}$) obtained from Band 3 ToAs (blue points), and using the wideband CC (magenta points) and CP (green points) methods on the Band 3+5 combination ToAs by implementing wideband timing technique in \texttt{tempo2}, are plotted against the corresponding epochs. Pulsar names and their respective post-fit weighted RMS residuals for Band 3, CC, and CP methods (Band 3+5) are mentioned at the bottom of the respective panels. The vertical axes in each panel are scaled differently for 100 MHz (left axis) and 200 MHz (right axis) epochs and a vertical dashed line is added such that the residual variations are clearly visible. Epochs are depicted in terms of the Modified Julian Date on the consolidated horizontal axes.}
    \label{fig:res_comp_plot}
\end{figure*}

A consolidated wideband timing residual plot obtained from Band 3, CC, and CP ToAs for all the 14 InPTA DR1 pulsars is shown in figure \ref{fig:res_comp_plot}. The timing procedure that we followed is the same as that described in sections \ref{timingprocedure} and \ref{timingJ1909-3744}. The \texttt{DMEFAC} and \texttt{T2EFAC} values are estimated separately for $100$ MHz and $200$ MHz BW data (as they have different sensitivities) for each of the Band 3, CC, and CP ToAs. We then incorporate the \texttt{T2EFAC} and \texttt{DMEFAC} values along with combining the 100 MHz and 200 MHz BW data ToAs to generate a global timing solution. The details of timing parameter estimates obtained after wideband timing using Band 3, CC, and CP method ToAs for all 14 pulsars are mentioned in table \ref{tab:3}, where the fit parameters are chosen as per InPTA DR1 narrowband timing \citep{TarafdarNobleson+2022}. We find that the precision of the fitted parameters are improved when the timing is done on Band 3+5 data using both CC and CP ToAs for most of the pulsars.

\begin{sidewaystable*}[!ht]
\vspace{18cm}
\begin{adjustbox}{width=\columnwidth,center}
\scriptsize
\centering
\begin{tabular}{ccccccccccccccccc} \hline \\ 
\multirow{10}{*}{\textbf{Pulsar}} &\multirow{10}{*}{\textbf{Method}} &\multicolumn{12}{c}{\multirow{2}{*}{\textbf{\large{Timing Parameters}}}} \\ \\ \hline
& & & & & & & & & & & & & & & \\
& &\multicolumn{2}{c}{\multirow{2}{*}{\textbf{RA} (hh:mm:ss)}} &\multicolumn{2}{c}{\multirow{2}{*}{\textbf{DEC} ($\deg$:mm:ss)}} &\multicolumn{2}{c}{\multirow{2}{*}{\textbf{PMRA} (mas/yr)}} &\multicolumn{2}{c}{\multirow{2}{*}{\textbf{A1} (lt s)}} &\multicolumn{2}{c}{\multirow{2}{*}{\textbf{F0} ($\mathrm{s}^{-1}$)}} &\multicolumn{2}{c}{\multirow{2}{*}{\textbf{F1} ($\mathrm{s}^{-2}$)}} &\multicolumn{2}{c}{\multirow{2}{*}{\textbf{PB} (s)}} \\
& & & & & & & & & & & & & & & \\
& &Value &Error ($\mathrm{s}$) &Value &Error ($\mathrm{s}$) &Value &Error ((mas/yr)) &Value &Error ($\mathrm{lt} \ \mathrm{s}$) &Value &Error ($\mathrm{s}^{-1}$) &Value &Error ($\mathrm{s}^{-2}$) &Value &Error ($\mathrm{s}$) \\ \\ \hline \\
\multirow{3}{*}{J0437$-$4715} &Band 3 &$4:37:16.0432$ &$5.9\times10^{-4}$ &$-47:15:09.991$ &$9.5\times10^{-3}$ &------ &------ &$3.36673$ &$1.2\times10^{-5}$ & $173.6879451409$ &$3.7\times10^{-10}$ &------ &------ &$5.741045808$ &$5.0\times10^{-9}$ \\
&CC &$4:37:16.04316$ &$8.3\times10^{-5}$ &$-47:15:09.991$ &$1.1\times10^{-3}$ &------ &------ &$3.366759$ &$1.6\times10^{-6}$ &$173.68794514023$ &$4.2\times10^{-11}$ &------ &------ &$5.7410458046$ &$8.1\times10^{-10}$ \\
&CP &$4:37:16.04316$ &$9.1\times10^{-5}$ &$-47:15:09.991$ &$1.2\times10^{-3}$ &------ &------ &$3.366759$ &$1.8\times10^{-6}$ &$173.68794514020$ &$4.5\times10^{-11}$ &------ &------ &$5.74104580486$ &$8.8\times10^{-10}$ \\ \\ 
J0613$-$0200 &------ &------ &------ &------ &------ &------ &------ &------ &------ &------ &------ &------ &------ &------ &------ \\ \\ 
J0751$+$1807 &------ &------ &------ &------ &------ &------ &------ &------ &------ &------ &------ &------ &------ &------ &------ \\ \\ 
\multirow{3}{*}{J1012$+$5307} &Band 3 &------ &------ &------ &------ &------ &------ &------ &------ &$190.26783422728$ &$1.0\times10^{-11}$ &------ &------ &------ &------ \\
&CC &------ &------ &------ &------ &------ &------ &------ &------ &$190.267834227285$ &$4.8\times10^{-12}$ &------ &------ &------ &------ \\
&CP &------ &------ &------ &------ &------ &------ &------ &------ &$190.267834227289$ &$4.8\times10^{-12}$ &------ &------ &------ &------ \\ \\ 
J1022+1001 &------ &------ &------ &------ &------ &------ &------ &------ &------ &------ &------ &------ &------ &------ &------ \\ \\
\multirow{3}{*}{J1600$-$3053} &Band 3 &------ &------ &------ &------ &------ &------ &------ &------ &$277.937706735932$ &$6.8\times10^{-11}$ &------ &------ &------ &------ \\
&CC &------ &------ &------ &------ &------ &------ &------ &------ &$277.9377067359737$ &$8.9\times10^{-12}$ &------ &------ &------ &------ \\
&CP &------ &------ &------ &------ &------ &------ &------ &------ &$277.937706735978$ &$8.8\times10^{-12}$ &------ &------ &------ &------ \\ \\ 
\multirow{3}{*}{J1643$-$1224} &Band 3 &------ &------ &------ &------ & $5.5$ & $1.0$ & $25.072598$ & $1.7\times10^{-6}$ & $216.37333684399$ & $1.2\times 10^{-11}$ & $-8.599\times10^{-16}$ & $8.2\times10^{-19}$ & $147.01739775$ & $4.0\times10^{-8}$ \\
&CC &------ &------ &------ &------ & $7.1$ & $0.4$ & $25.0725981$ & $8.9\times10^{-7}$ & $216.373336843944$ & $6.3\times10^{-12}$ & $-8.637\times10^{-16}$ & $3.2\times10^{-19}$ & $147.01739786$ & $2.1\times10^{-8}$ \\
&CP &------ &------ &------ &------ & $7.2$ & $0.4$ & $25.0725978$ & $9.3\times10^{-7}$ & $216.373336843935$ & $6.5\times10^{-12}$ & $-8.635\times10^{-16}$ & $3.3\times10^{-19}$ & $147.01739786$ & $2.3\times10^{-8}$ \\ \\ 
\multirow{3}{*}{J1713$+$0747} &Band 3 &------ &------ &------ &------ &------ &------ & $32.3424310$ & $5.7\times10^{-7}$ &------ &------ &------ &------ & $67.825130884$ & $2.5\times10^{-9}$ \\
&CC &------ &------ &------ &------ &------ &------ & $32.3424298$ & $2.1\times10^{-7}$ &------ &------ &------ &------ & $67.8251308817$ & $8.9\times10^{-10}$ \\
&CP &------ &------ &------ &------ &------ &------ & $32.3424297$ & $2.5\times10^{-7}$ &------ &------ &------ &------ & $67.825130881$ & $1.0\times10^{-9}$ \\ \\ 
\multirow{3}{*}{J1744$-$1134} &Band 3 &------ &------ &------ &------ &------ &------ &------ &------ &$245.42611950378$ &$3.0\times10^{-11}$ &------ &------ &------ &------ \\
&CC &------ &------ &------ &------ &------ &------ &------ &------ &$245.42611950381$ &$2.6\times10^{-11}$ &------ &------ &------ &------ \\
&CP &------ &------ &------ &------ &------ &------ &------ &------ &$245.42611950381$ &$1.6\times10^{-11}$ &------ &------ &------ &------ \\ \\ 
\multirow{3}{*}{J1857$+$0943} &Band 3 &------ &------ &------ &------ &------ &------ &------ &------ & $186.49407816357$ & $1.5\times10^{-11}$ & $-6.22\times10^{-16}$ & $1.2\times10^{-18}$ &------ &------ \\
&CC &------ &------ &------ &------ &------ &------ &------ &------ & $186.494078163546$ & $3.6\times10^{-12}$ & $-6.205\times10^{-16}$ & $1.6\times10^{-19}$ &------ &------ \\
&CP &------ &------ &------ &------ &------ &------ &------ &------ & $186.494078163547$ & $2.8\times10^{-12}$ & $-6.204\times10^{-16}$ & $1.4\times10^{-19}$ &------ &------ \\ \\ 
\multirow{3}{*}{J1909$-$3744} &Band 3 &------ &------ &------ &------ &------ &------ &------ &------ &$339.31568666040$ &$1.0\times10^{-11}$ &$-1.6142\times10^{-15}$ &$4.4\times10^{-19}$ &------ &------ \\
&CC &------ &------ &------ &------ &------ &------ &------ &------ &$339.315686660410$ &$2.7\times10^{-12}$ &$-1.6149\times10^{-15}$ &$1.7\times10^{-19}$ &------ &------ \\
&CP &------ &------ &------ &------ &------ &------ &------ &------ &$339.315686660412$ &$2.3\times10^{-12}$ &$-1.6152\times10^{-15}$ &$1.4\times10^{-19}$ &------ &------ \\ \\ 
\multirow{3}{*}{J1939$+$2134} &Band 3 &------ &------ &------ &------ &------ &------ &------ &------ & $641.92820961498$ & $1.2\times10^{-11}$ & $-4.33096\times10^{-14}$ & $7.5\times10^{-19}$ &------ &------ \\
&CC &------ &------ &------ &------ &------ &------ &------ &------ & $641.928209615009$ & $6.7\times10^{-12}$ & $-4.33091\times10^{-14}$ & $4.1\times10^{-19}$ &------ &------ \\
&CP &------ &------ &------ &------ &------ &------ &------ &------ & $641.928209615108$ & $6.7\times10^{-12}$ & $-4.33058\times10^{-14}$ & $2.6\times10^{-19}$ &------ &------ \\ \\ 
J2124-3358 &------ &------ &------ &------ &------ &------ &------ &------ &------ &------ &------ &------ &------ &------ &------ \\ \\ 
\multirow{3}{*}{J2145$-$0750} &Band 3 &------ &------ &------ &------ &------ &------ &------ &------ &$62.295887797432$ &$1.7\times10^{-12}$ &$-1.155\times10^{-16}$ &$1.0\times10^{-19}$ &$6.8389026151$ &$1.0\times10^{-10}$ \\
&CC &------ &------ &------ &------ &------ &------ &------ &------ &$62.2958877974360$ &$9.1\times10^{-13}$ &$-1.1565\times10^{-16}$ &$4.5\times10^{-20}$ &$6.838902615$ &$5.5\times10^{-11}$ \\
&CP &------ &------ &------ &------ &------ &------ &------ &------ &$62.29588779744$ &$1.0\times10^{-12}$ &$-1.1547\times10^{-16}$ &$4.2\times10^{-20}$ &$6.8389026152$ &$4.4\times10^{-11}$ \\ \\ \hline \\
\end{tabular}
\end{adjustbox}
\caption{Table of timing parameters for 14 InPTA DR1 pulsars. 
The first column lists the pulsar names. 
The second column lists the methodology used to obtain ToAs which are then used for estimating the timing parameters in separate rows, namely the Band 3, CC, and CP methods. 
Columns three to nine represent various fitted pulsar timing parameters, their units, and their uncertainties. 
The choice of the fitted parameters for each pulsar is consistent with the timing analysis of \protect\cite{TarafdarNobleson+2022}, where no timing parameters are fit for PSRs J0613$-$0200, J0751+1807, J1022+1001, and J2124$-$3358.}
\label{tab:3}
\end{sidewaystable*}

\section{Summary and Conclusions}
\label{sec:summary}

\begin{table*}
\centering

\begin{tabular}{c}
\hline \\
\textbf{Table of wRMS and $\chi^2$} \\
[0.2cm]
\hline
\begin{tabular}{ccccccc}\\
\textbf{Pulsar Name}        & \textbf{Method} & \textbf{ToA $\chi^2$} & \textbf{DM $\chi^2$} & \textbf{DOF} & \textbf{Total red. $\chi^2$} & \textbf{wRMS} \\\\
\hline \\
\multirow{3}{*}{J0437$-$4715} & B3              & 0.087             & 10.196           & 9             & 1.143                    & 0.201                 \\
                            & CC              & 0.631             & 9.777            & 9             & 1.156                    & 0.217                 \\
                            & CP              & 0.585             & 8.928            & 9             & 1.057                    & 0.225                 \\ \\
\multirow{3}{*}{J0613$-$0200} & B3              & 7.034             & 17.621           & 21            & 1.174                    & 0.963                 \\
                            & CC              & 2.729             & 31.737           & 21            & 1.641                    & 0.555                 \\
                            & CP              & 7.271             & 21.607           & 21            & 1.375                    & 0.972                 \\ \\
\multirow{3}{*}{J0751$+$1807} & B3              & 0.845             & 22.384           & 23            & 1.01                     & 0.654                 \\
                            & CC              & 0.785             & 27.978           & 23            & 1.251                    & 0.348                 \\
                            & CP              & 1.582             & 35.636           & 23            & 1.618                    & 0.302                 \\ \\
\multirow{3}{*}{J1012$+$5307} & B3              & 6.088             & 28.914           & 22            & 1.591                    & 0.77                  \\
                            & CC              & 6.631             & 24.177           & 22            & 1.4                      & 0.53                  \\
                            & CP              & 2.558             & 21.554           & 22            & 1.096                    & 0.306                 \\ \\
\multirow{3}{*}{J1022$+$1001} & B3              & 0.398             & 21.283           & 24            & 0.903                    & 0.264                 \\
                            & CC              & 0.448             & 20.798           & 24            & 0.885                    & 0.214                 \\
                            & CP              & 0.734             & 32.475           & 24            & 1.384                    & 0.286                 \\ \\
\multirow{3}{*}{J1600$-$3053} & B3              & 1.72              & 19.029           & 21            & 0.988                    & 1.018                 \\
                            & CC              & 9.953             & 16.848           & 21            & 1.276                    & 1.804                 \\
                            & CP              & 13.781            & 13.766           & 21            & 1.312                    & 2.178                 \\ \\
\multirow{3}{*}{J1643$-$1224} & B3              & 9.984             & 164.109          & 63            & 2.763                    & 0.869                 \\
                            & CC              & 4.317             & 63.795           & 63            & 1.081                    & 0.257                 \\
                            & CP              & 3.799             & 63.037           & 63            & 1.061                    & 0.269                 \\ \\
\multirow{3}{*}{J1713$+$0747} & B3              & 5.203             & 49.358           & 43            & 1.269                    & 0.399                 \\
                            & CC              & 18.088            & 48.758           & 43            & 1.555                    & 0.417                 \\
                            & CP              & 25.567            & 111.641          & 43            & 3.191                    & 0.516                 \\ \\
\multirow{3}{*}{J1744$-$1134} & B3              & 3.859             & 11.284           & 14            & 1.082                    & 0.49                  \\
                            & CC              & 3.413             & 11.203           & 14            & 1.044                    & 0.434                 \\
                            & CP              & 2.348             & 10.309           & 14            & 0.904                    & 0.255                 \\ \\
\multirow{3}{*}{J1857$+$0943} & B3              & 1.315             & 31.023           & 34            & 0.951                    & 0.323                 \\
                            & CC              & 6.395             & 38.517           & 34            & 1.321                    & 0.352                 \\
                            & CP              & 11.967            & 33.109           & 34            & 1.326                    & 0.589                 \\ \\
\multirow{3}{*}{J1909$-$3744} & B3              & 5.944             & 50.301           & 56            & 1.004                    & 0.235                 \\
                            & CC              & 11.395            & 45.588           & 56            & 1.018                    & 0.326                 \\
                            & CP              & 19.304            & 35.852           & 56            & 0.985                    & 0.471                 \\ \\
\multirow{3}{*}{J1939$+$2134} & B3              & 2.827             & 61.999           & 63            & 1.029                    & 0.148                 \\
                            & CC              & 9.856             & 69.861           & 63            & 1.265                    & 0.253                 \\
                            & CP              & 23.987            & 74.188           & 63            & 1.558                    & 0.329                 \\ \\
\multirow{3}{*}{J2124$-$3358} & B3              & 4.094             & 63.051           & 39            & 1.722                    & 0.515                 \\
                            & CC              & 4.697             & 67.579           & 39            & 1.853                    & 0.324                 \\
                            & CP              & 5.501             & 73.437           & 39            & 2.024                    & 0.54                  \\ \\
\multirow{3}{*}{J2145$-$0750} & B3              & 2.284             & 40.514           & 44            & 0.973                    & 0.231                 \\
                            & CC              & 11.141            & 60.259           & 44            & 1.623                    & 0.335                 \\
                            & CP              & 19.252            & 92.857           & 44            & 2.548                    & 0.472 \\
\hline 
\end{tabular}
\end{tabular}
\caption{Table of reduced chi-squares of 14 InPTA DR1 pulsars. 
The first column lists the pulsar names. 
The second column lists the methodology used to obtain ToAs  and DM, namely the Band 3, CC, and CP methods. 
Columns three to seven represent ToA component of the chi-square, DM component of the chi-square, degrees of freedom (DOF), the total reduced chi-square and the wRMS of the timing residuals. }
\label{tab:4}
\end{table*}

In this work, we have developed two independent novel techniques, namely the Combined Portrait (CP) and Combined Chi-squared (CC) methods, to combine data simultaneously recorded in two non-contiguous frequency bands within the paradigm of wideband technique \citep{PennucciDemorestRansom2014,Pennucci2019} to obtain a single DM and ToA per epoch encapsulating information contained in both the bands. In the CP method, we create an auxiliary dataset by combining the data of two frequency bands to create a single 2-dimensional analytic template containing the information on pulse profile evolution with frequency. This template is then used for cross-correlation with other epochs to obtain wideband DMs and ToAs.
In the CC method, we create separate 2-dimensional analytic templates for both the bands, and these are integrated within a combined Fourier-domain $\chi^2$-statistic and perform a global fit over the whole frequency space to generate a single wideband DM and ToA per epoch. We have applied these two techniques to 14 millisecond pulsars observed under the InPTA campaign using uGMRT in Band 3 and Band 5 frequency bands simultaneously, and they are included in the first data release of the InPTA \citep{TarafdarNobleson+2022}. 

We obtained high-precision DMs and ToAs for Band 3+5 data using these techniques. We observe that combining the data having 100 MHz bandwidth in each band showed consistent improvement in DM and ToA precision for all 14 pulsars and that both CC and CP are performing equally well. However, the combination of data having 200 MHz bandwidth in each band shows inconsistencies using the CP method. This is due to the reduction in the number of phase bins in Band 3 which is essential to combine it with Band 5 data to create a single analytic template of Band 3+5. Another caveat of the CP method is the band gap of $\sim 760$ MHz, which needs to be interpolated over, between two bands leading to probable imperfections in the modeling of profile evolution with frequency across the bands. The combination of data having 200 MHz bandwidth in each band using the CC method shows much higher improvement in DM and ToA precision than the CP method and Band 3 alone. We plan to extend these techniques further to combine simultaneously recorded data of multiple non-contiguous bands in future work.

We have also incorporated the wideband likelihood in \tempotwo{} using \libstempo{} for the first time. We perform the wideband timing analysis on ToAs obtained from the CC and CP methods along with Band 3 ToAs for comparison. We achieved the weighted RMS ToA residuals in the range of 214 ns to 1.8 $\mu$s for ToAs obtained from the CC method, while in the range of 225 ns to 2.1 $\mu$s for ToAs obtained from the CP method for the whole spectrum of InPTA DR1 pulsars. We observe an improvement in the precision of fitted timing parameters with Band 3+5 combination compared to Band 3 alone for all pulsars. Since we are combining data of multiple frequency bands, we may require frequency-dependent parameters to obtain a better fit for our timing solutions. This will be explored in future work.

We observe that the DM chi-square, obtained from the DM part of the likelihood, is larger than the ToA chi-square (see Table \ref{tab:4}).  We suspect that this could be related to the way we estimate DMEFAC and T2EFAC parameters. We plan to investigate this further in future work where we will apply Bayesian methods to estimate optimum DMEFAC and T2EFAC parameter values.

\section{Discussion and Future directions}
\label{sec:discussion}

The extension of the wideband technique to multiple non-contiguous frequency 
bands demonstrated in this work, is likely to be useful in largely 
removing chromatic noise sources, such as variations in the pulse profile, 
DM and scattering, in precision timing experiments like pulsar timing arrays. 
This technique not only improves the ToA precision significantly by accumulating 
the signal over the entire frequency range of combined bands, it also takes care of DM noise across the bands by incorporating 
DM-chromatic noise measurements in the timing likelihood naturally (see Appendix \ref{appB}). This restricts 
the noise analysis of PTA data to just the time-independent and time-correlated achromatic and scattering
noise sources, greatly simplifying and constraining these noise models. This has 
implications both for the computational needs as well as the sensitivity of PTA data for a 
GW search. Other precision timing experiments targeted at  measuring timing noise, 
parameters of relativistic binary systems and tests of General Theory 
of Relativity are also likely to benefit from this extension of the standard wideband technique. With 
large upcoming and future telescopes, such as the
SKA \citep{KramerStappers+2015,JanssenHobbs+2015} and 
DSA \citep{HallinanRavi+2019}, likely to employ simultaneous observations over multiple 
bands with frequency coverage as large as 5 GHz, we expect this extended technique or its variants to be widely used in the future.

\section*{Software}
\texttt{RFIclean} \citep{MaanvanLeeuwenVohl2020}, 
\texttt{DSPSR} \citep{vanStratenBailes2011}, 
\texttt{PSRCHIVE} \citep{HotanvanStratenManchester2004}, 
\texttt{pinta} \citep{SusobhananMaan+2021}, 
\pp{} \citep{PennucciDemorestRansom2014,Pennucci2019},
\texttt{tempo2} \citep{HobbsEdwardsManchester2006,EdwardsHobbsManchester2006}, 
\texttt{libstempo} \citep{Vallisneri2020}, 
\texttt{tempo} \citep{NiceDemorest+2015},
\texttt{numpy} \citep{HarrisMillman+2020},
\texttt{scipy} \citep{PauliGommers+2020},
\texttt{matplotlib} \citep{Hunter2007}


\section*{Acknowledgements}
We thank the staff of the GMRT who made our observations possible. 
GMRT is operated by the National Centre for Radio Astrophysics of the Tata Institute of Fundamental Research. AKP is supported by CSIR fellowship Grant number $09/0079(15784)/2022$-EMR-I. BCJ acknowledges support from Raja Ramanna Chair (Track - I) grant from the Department of Atomic Energy, Government of India. KN is supported by the Birla Institute of Technology and Science Institute fellowship. AS is supported by the NANOGrav NSF Physics Frontiers Center (awards 1430284 and 2020265). DD acknowledges the support from the Department of Atomic Energy, Government of India through `Apex Project - Advance Research and Education in Mathematical Sciences at IMSc'. MB acknowledges the support from the Department of Atomic Energy, Government of India through `Apex Project - Advance Research and Education in Mathematical Sciences at IMSc'. YG and BCJ acknowledges support from the Department of Atomic Energy, Government of India, under project number 12-R\&D-TFR-5.02-0700. TK is partially supported by the JSPS Overseas Challenge Program for Young Researchers. AmS is supported by CSIR fellowship Grant number $09/1001(12656)/2021$-EMR-I
and DST-ICPS T-641. KT is partially supported by JSPS KAKENHI Grant Numbers 20H00180, 21H01130, and 21H04467 and the ISM Cooperative Research Program (2023-ISMCRP-2046). We thank Scott Ransom for his suggestions that improved the manuscript.

\section*{Data Availability}
The python scripts used for the analysis are available in \url{https://github.com/AvinashKumarPaladi/Multiband-extension-of-Wideband-Timing-Technique}. The data underlying this article will be shared on reasonable request to the corresponding author. 



\bibliographystyle{mnras}
\bibliography{inpta-wideband-2} 



\onecolumn
\appendix

\section{DM and ToA Uncertainty Comparison Plots}\label{appA}

The DM and ToA uncertainty ($\sigma_{\text{DM}}$ and $\sigma_{\text{ToA}}$) comparison between Band 3 and Band 3+5 data -- CC and CP methods is presented here for all the 14 InPTA DR1 pulsars. Since the sensitivity of both bandwidths is different, it is not visually feasible to plot all of them on the same scale, hence we have shown them in different panels. 

\begin{figure}[ht!]
    \begin{center}
    \includegraphics[scale=1.1, width=0.9\linewidth]{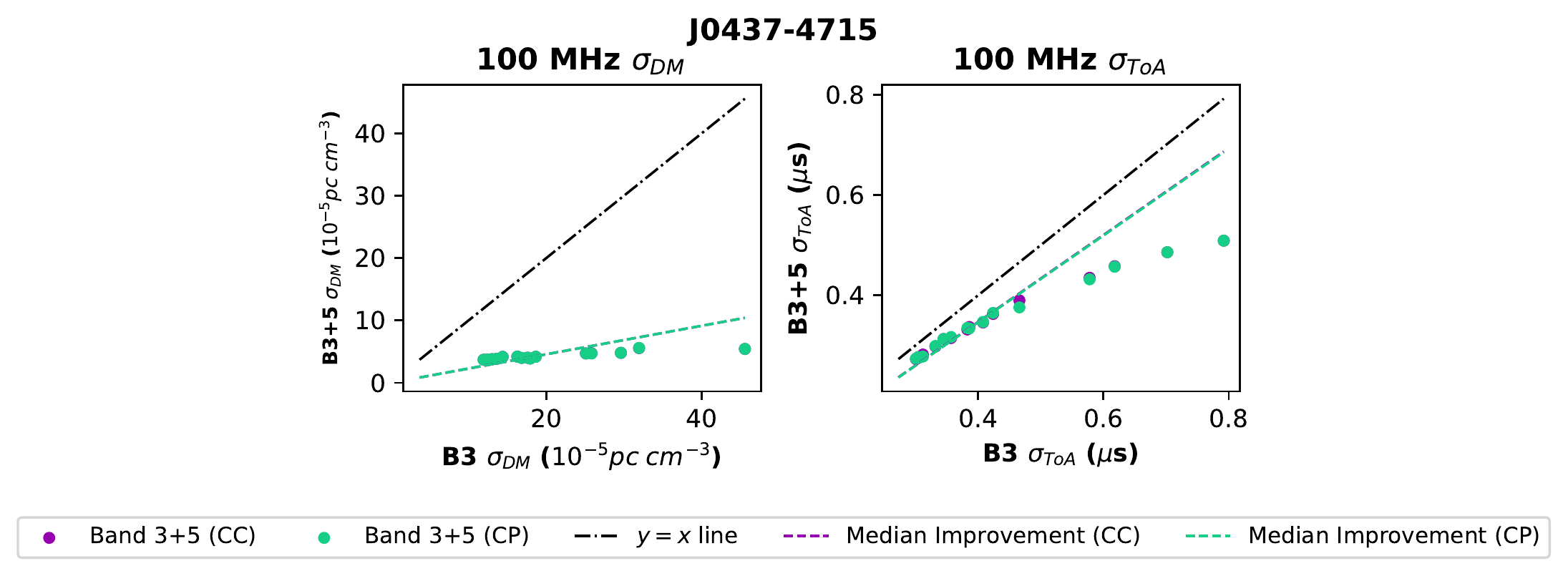}
    \caption{A comparison of Wideband DM (left panel) and ToA (right panel) uncertainties obtained from the CC and CP methods for Band 3+5 data is shown against those obtained from the traditional Band 3 data for PSR J0437$-$4715. This pulsar has only 100MHz InPTA data before Cycle 40 of the uGMRT. The Band 3+5 uncertainties estimated with the CC (magenta points) and CP (green points) methods are shown on the vertical axis and Band 3 uncertainties on the horizontal axis (in units of $10^{-5}\;\text{pc}\;\text{cm}^{-3}$ for DM uncertainties, and in units of $\mu\text{s}$ for ToA uncertainties). The diagonal dashed-dotted line shows the $y=x$ while the magenta and green dashed lines indicate median DM (left panel) and ToA (right panel) uncertainties for CC and CP method respectively. Points lying below the diagonal line indicate that the obtained precision with Band 3+5 combination is better than Band 3 (single-band) results and vice versa. Hence, we obtain an overall increase in the DM and ToA precisions from both CP and CC methods for PSR J0437$-$4715.}
    \end{center}
    \label{fig:A1}
\end{figure}

\begin{figure}[ht!]
    \includegraphics[scale=1.1, width=\linewidth]{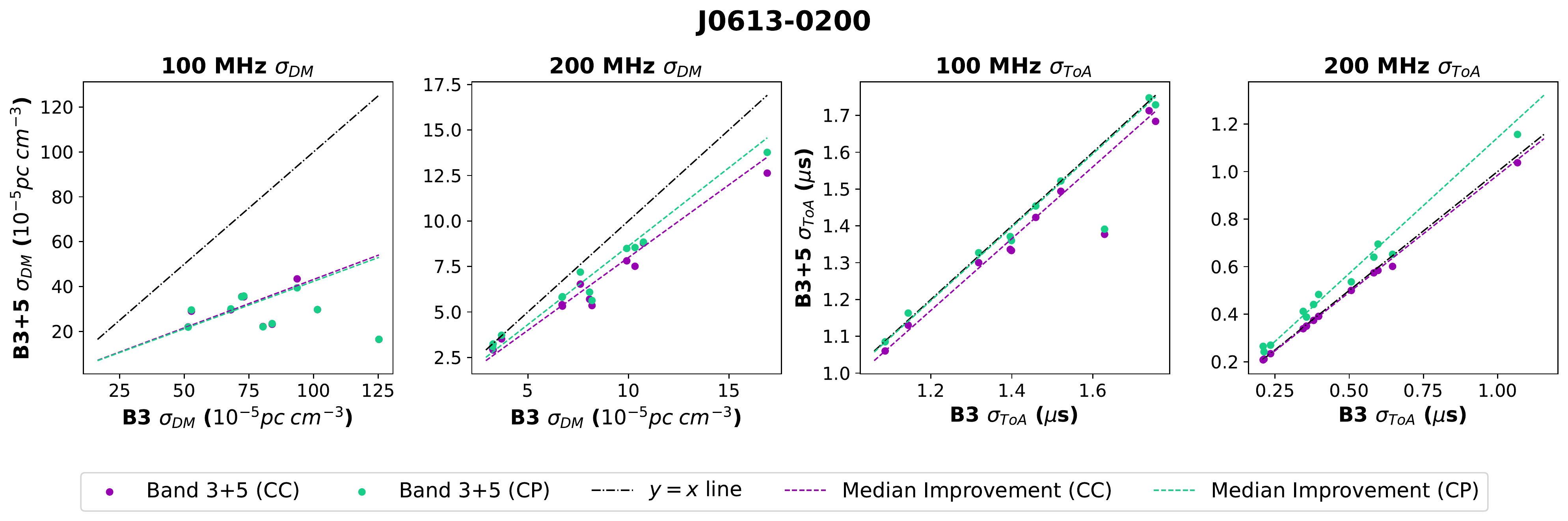}
    \caption{A comparison of Wideband DM (left panel) and ToA (right panel) uncertainties obtained from the CC and CP methods for Band 3+5 data is shown against those obtained from the traditional Band 3 data for PSR J0613$-$0200. The results are shown for both 100 MHz and 200 MHz bandwidth data. The curves and colors displayed in the figure conform to the details mentioned in the legend at the bottom.}
    \label{fig:A2}
\end{figure}

\begin{figure}[ht!]
    \includegraphics[scale=1.1, width=\linewidth]{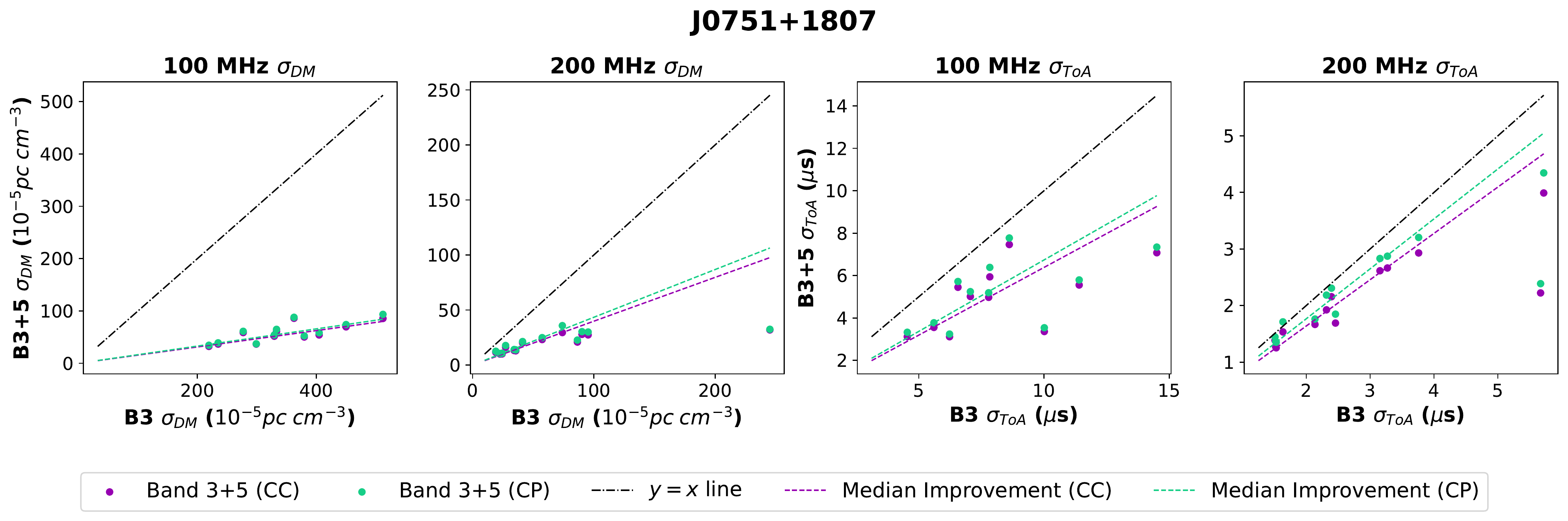}
    \caption{  A comparison of Wideband DM (left panel) and ToA (right panel) uncertainties obtained from the CC and CP methods for Band 3+5 data is shown against those obtained from the traditional Band 3 data for PSR J0751$+$1807. The results are shown for both 100  MHz and 200  MHz bandwidth data. The curves and colors displayed in the figure conform to the details mentioned in the legend at the bottom.}
    \label{fig:A3}
\end{figure}

\begin{figure}[ht!]
    \includegraphics[scale=1.1, width=\linewidth]{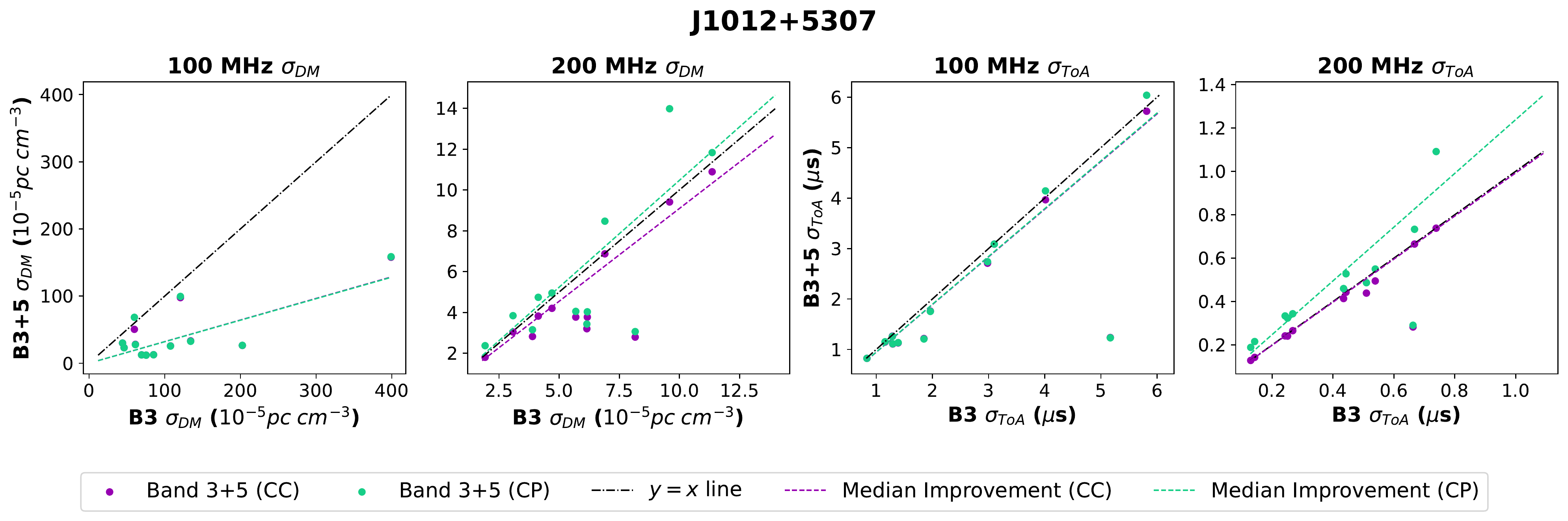}
    \caption{ A comparison of Wideband DM (left panel) and ToA (right panel) uncertainties obtained from the CC and CP methods for Band 3+5 data is shown against those obtained from the traditional Band 3 data for PSR J1012$+$5307. The results are shown for both 100 MHz and 200 MHz bandwidth data. The curves and colors displayed in the figure conform to the details mentioned in the legend at the bottom.}
    \label{fig:A4}
\end{figure}

\begin{figure}[ht!]
    \includegraphics[scale=1.1, width=\linewidth]{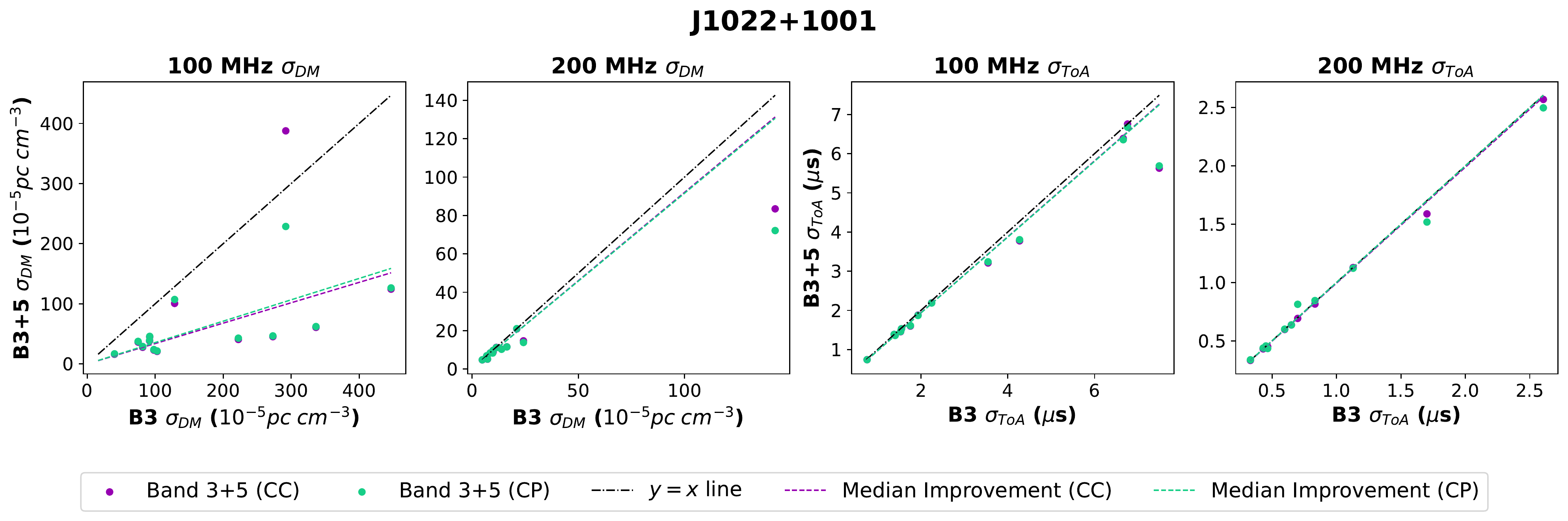}
    \caption{A comparison of Wideband DM (left panel) and ToA (right panel) uncertainties obtained from the CC and CP methods for Band 3+5 data is shown against those obtained from the traditional Band 3 data for PSR J1022$+$1001. The results are shown for both 100 MHz and 200 MHz bandwidth data. The curves and colors displayed in the figure conform to the details mentioned in the legend at the bottom.}
    \label{fig:A5}
\end{figure}

\begin{figure}[ht!]
    \includegraphics[scale=1.1, width=\linewidth]{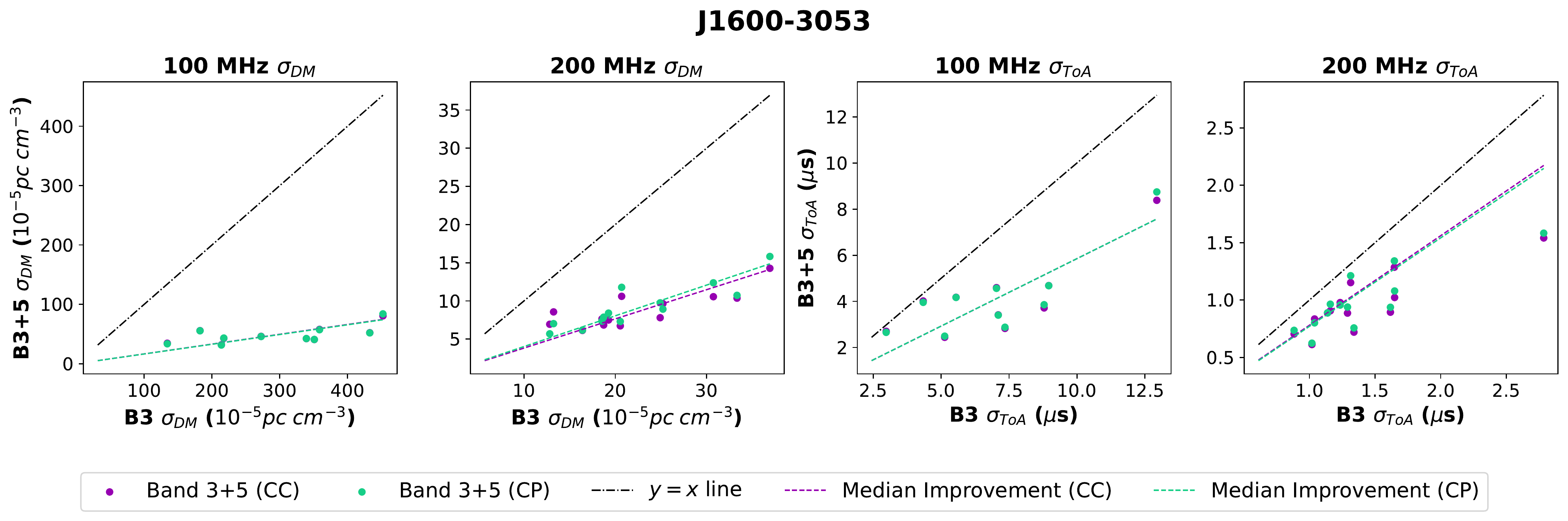}
    \caption{A comparison of Wideband DM (left panel) and ToA (right panel) uncertainties obtained from the CC and CP methods for Band 3+5 data is shown against those obtained from the traditional Band 3 data for PSR J1600$-$3053. The results are shown for both 100 MHz and 200 MHz bandwidth data. The curves and colors displayed in the figure conform to the details mentioned in the legend at the bottom.}
    \label{fig:A6}
\end{figure}

\begin{figure}[ht!]
    \includegraphics[scale=1.1, width=\linewidth]{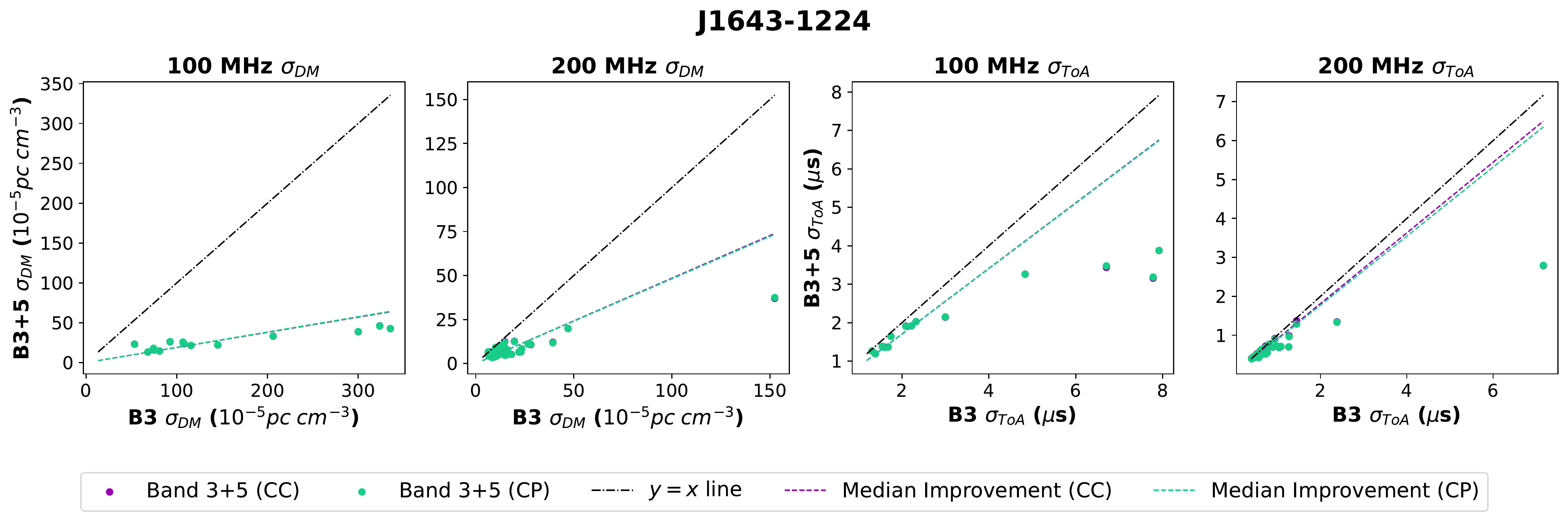}
    \caption{ A comparison of Wideband DM (left panel) and ToA (right panel) uncertainties obtained from the CC and CP methods for Band 3+5 data is shown against those obtained from the traditional Band 3 data for PSR J1643$-$1224. The results are shown for both 100 MHz and 200 MHz bandwidth data. The curves and colors displayed in the figure conform to the details mentioned in the legend at the bottom.}
    \label{fig:A7}
\end{figure}

\begin{figure}[ht!]
    \includegraphics[scale=1.1, width=\linewidth]{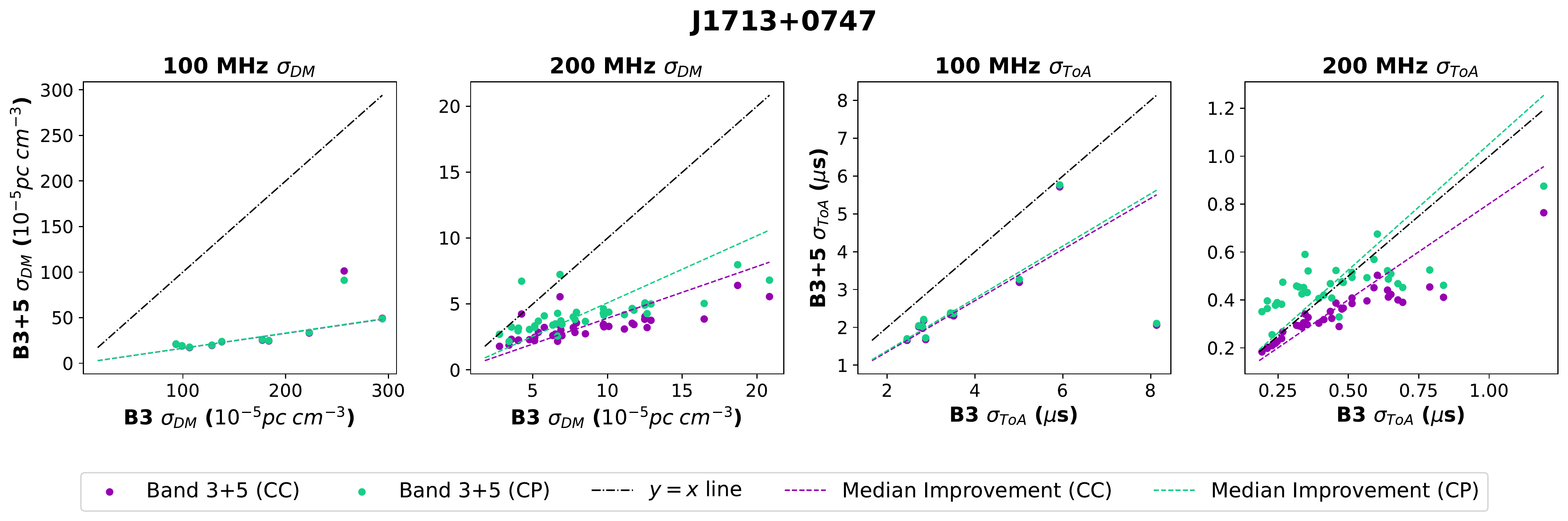}
    \caption{A comparison of Wideband DM (left panel) and ToA (right panel) uncertainties obtained from the CC and CP methods for Band 3+5 data is shown against those obtained from the traditional Band 3 data for PSR J1713$+$0747. The results are shown for both 100 MHz and 200 MHz bandwidth data. The curves and colors displayed in the figure conform to the details mentioned in the legend at the bottom.}
    \label{fig:A8}
\end{figure}

\begin{figure}[ht!]
    \begin{center}
    \includegraphics[scale=1.1, width=0.9\linewidth]{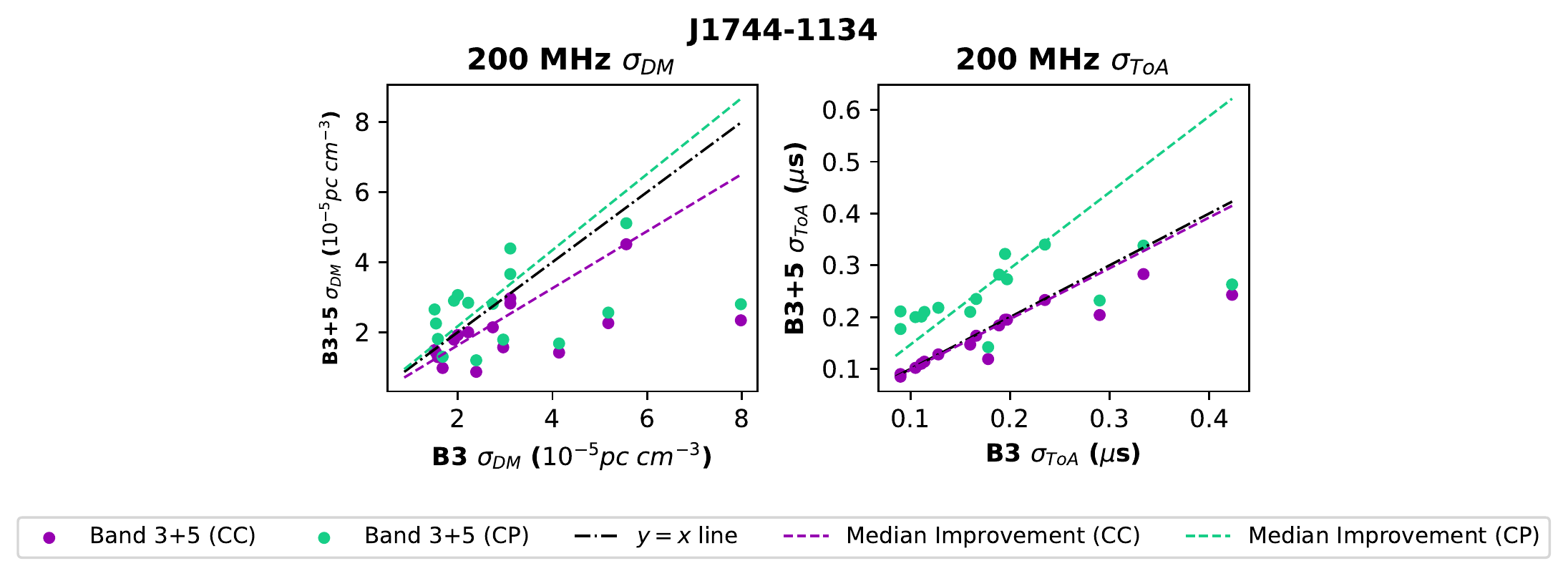}
    \caption{A comparison of Wideband DM (left panel) and ToA (right panel) uncertainties obtained from the CC and CP methods for Band 3+5 data is shown against those obtained from the traditional Band 3 data for PSR J1744$-$1134 with only 200 MHz data as it wasn't observed in earlier cycles. The curves and colors displayed in the figure conform to the details mentioned in the legend at the bottom.}
    \end{center}
    \label{fig:A9}
\end{figure}

\begin{figure}[ht!]
    \includegraphics[scale=1.1, width=\linewidth]{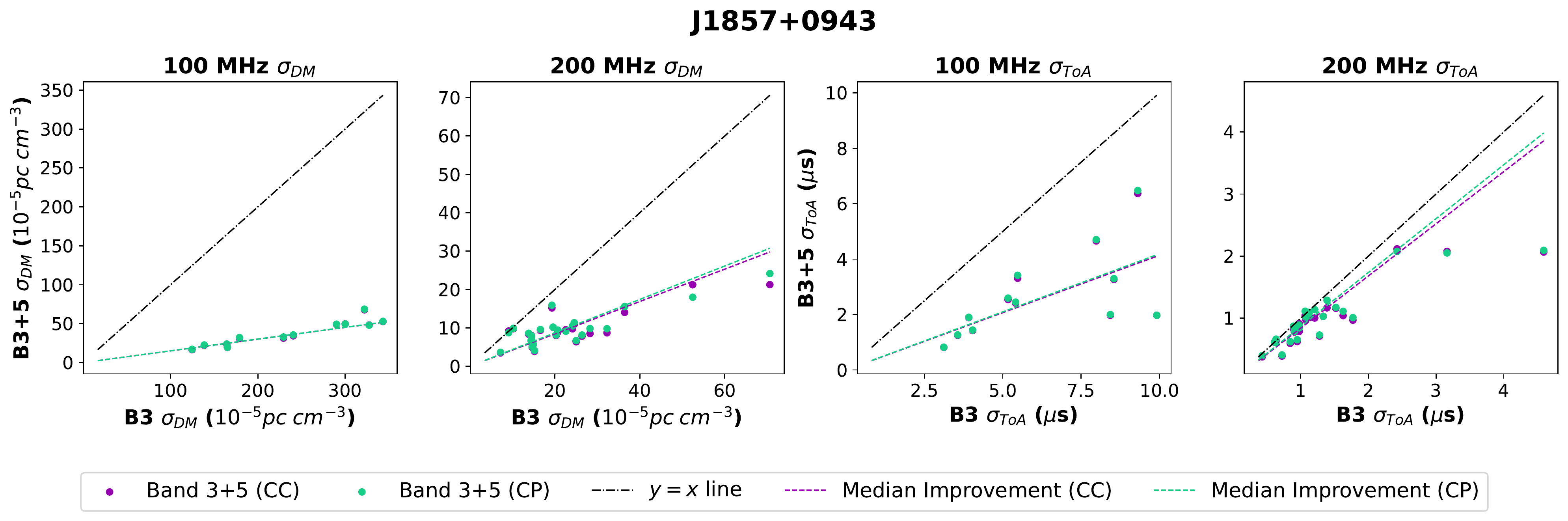}
    \caption{ A comparison of Wideband DM (left panel) and ToA (right panel) uncertainties obtained from the CC and CP methods for Band 3+5 data is shown against those obtained from the traditional Band 3 data for PSR J1857$+$0943. The results are shown for both 100 MHz and 200 MHz bandwidth data. The curves and colors displayed in the figure conform to the details mentioned in the legend at the bottom.
    }
    \label{fig:A10}
\end{figure}

\begin{figure}[ht!]
    \includegraphics[scale=1.1, width=\linewidth]{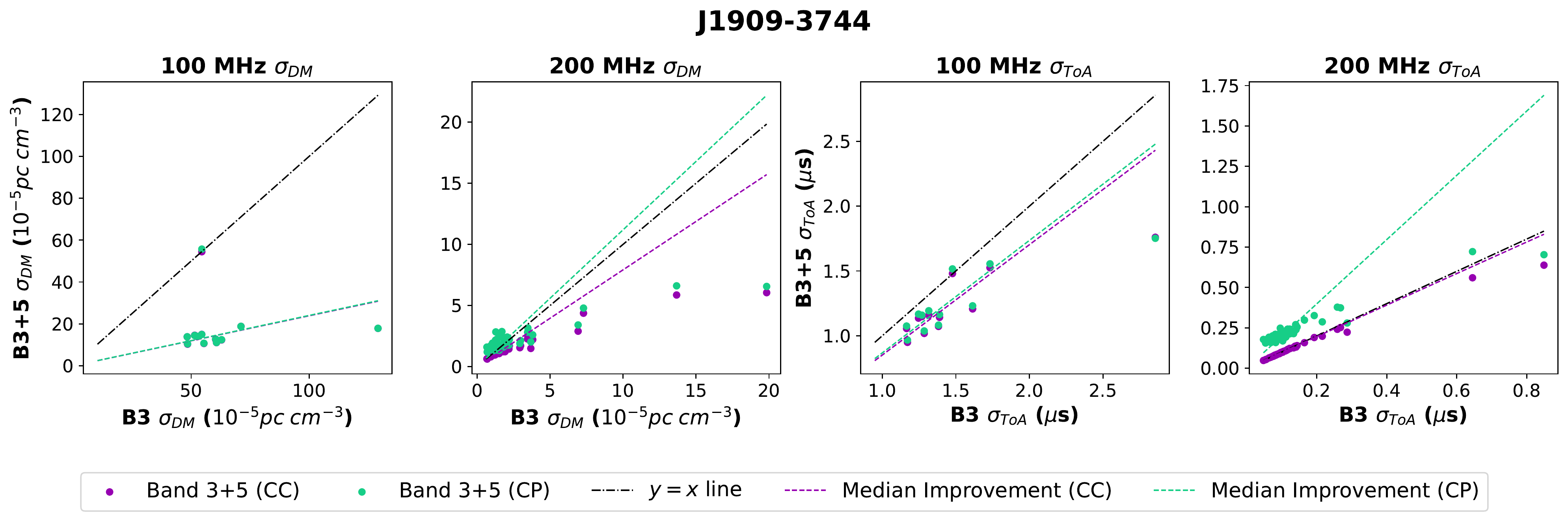}
    \caption{A comparison of Wideband DM (left panel) and ToA (right panel) uncertainties obtained from the CC and CP methods for Band 3+5 data is shown against those obtained from the traditional Band 3 data for PSR J1909$-$3744. The results are shown for both 100 MHz and 200 MHz bandwidth data. The curves and colors displayed in the figure conform to the details mentioned in the legend at the bottom.}
    \label{fig:A11}
\end{figure}

\begin{figure}[ht!]
    \includegraphics[scale=1.1, width=\linewidth]{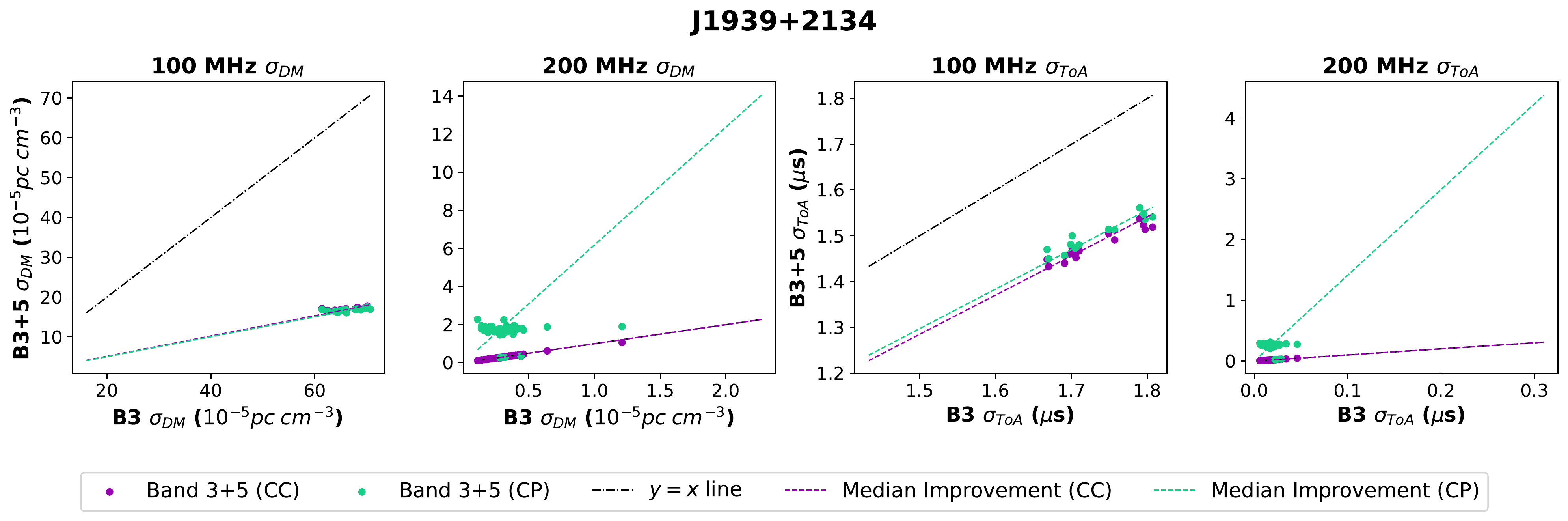}
    \caption{A comparison of Wideband DM (left panel) and ToA (right panel) uncertainties obtained from the CC and CP methods for Band 3+5 data is shown against those obtained from the traditional Band 3 data for PSR J1939$+$2134. The results are shown for both 100 MHz and 200 MHz bandwidth data. The curves and colors displayed in the figure conform to the details mentioned in the legend at the bottom.}
    \label{fig:A12}
\end{figure}

\begin{figure}[ht!]
    \includegraphics[scale=1.1, width=1.0\linewidth]{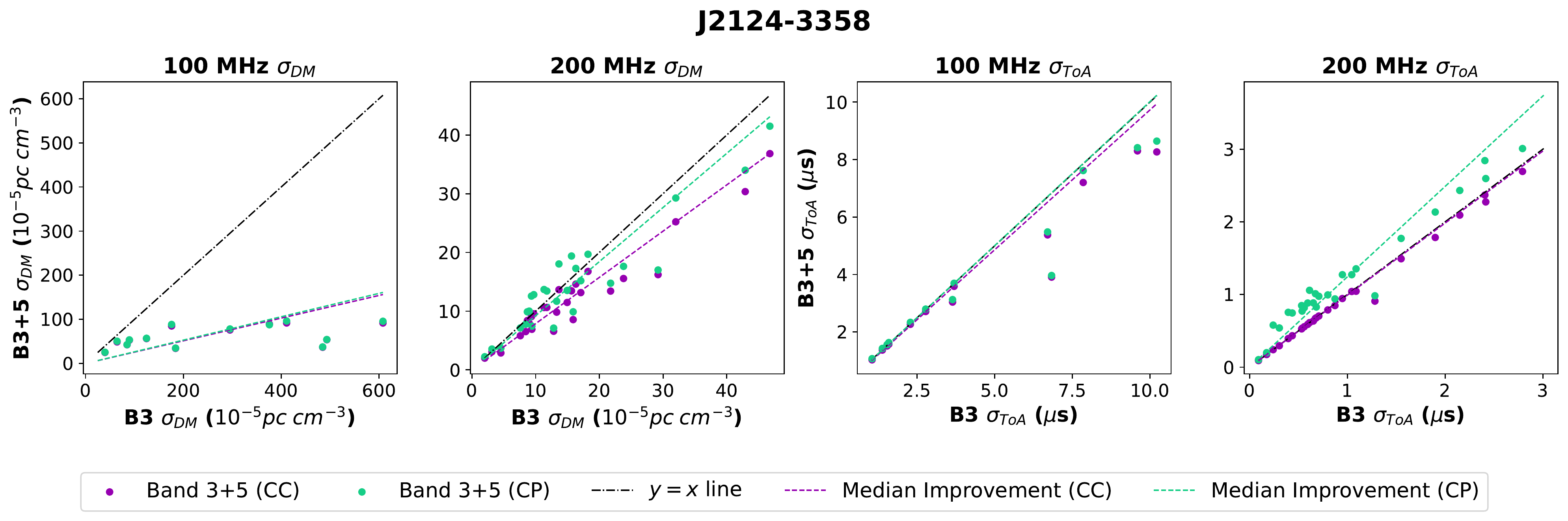}
    \caption{A comparison of Wideband DM (left panel) and ToA (right panel) uncertainties obtained from the CC and CP methods for Band 3+5 data is shown against those obtained from the traditional Band 3 data for PSR J2124$-$3358. The results are shown for both 100 MHz and 200 MHz bandwidth data. The curves and colors displayed in the figure conform to the details mentioned in the legend at the bottom.}
    \label{fig:A13}
\end{figure}

\begin{figure}[ht!]
    \includegraphics[scale=1.1, width=1.0\linewidth]{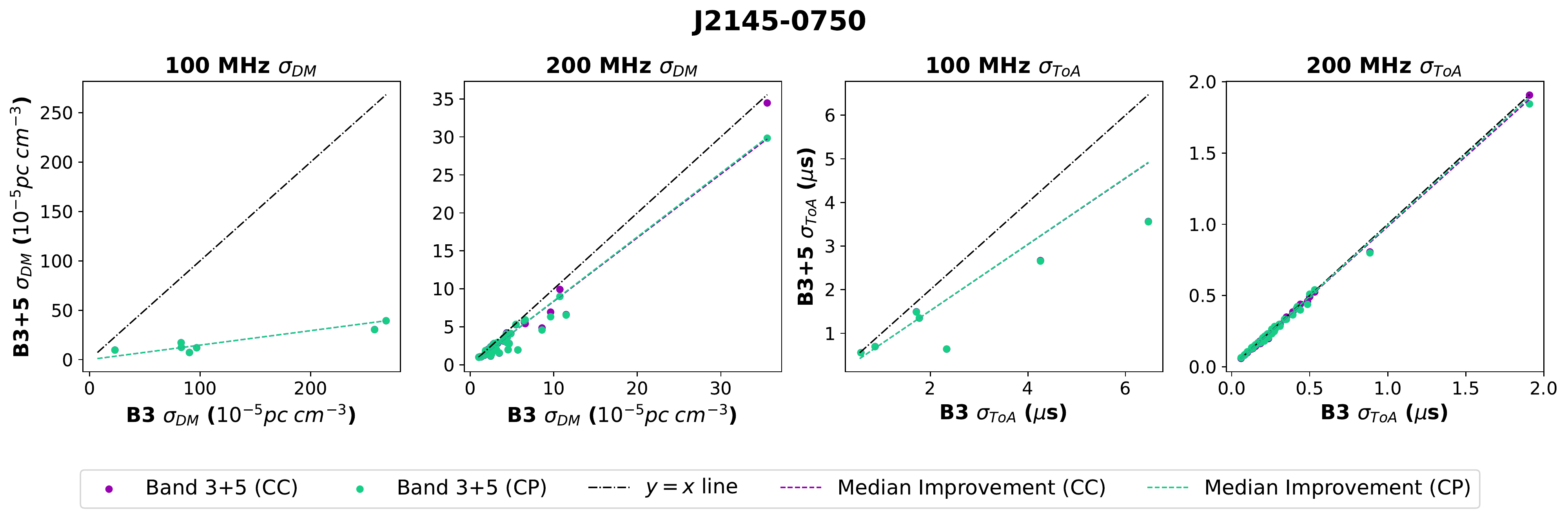}
    \caption{A comparison of Wideband DM (left panel) and ToA (right panel) uncertainties obtained from the CC and CP methods for Band 3+5 data is shown against those obtained from the traditional Band 3 data for PSR J2145$-$0750. The results are shown for both 100 MHz and 200 MHz bandwidth data. The curves and colors displayed in the figure conform to the details mentioned in the legend at the bottom.}
    \label{fig:A14}
\end{figure}

\newpage

\section{Implementing the wideband likelihood using \tempotwo{} and \libstempo{}} \label{appB}
The wideband timing residuals $\delta t$ can be modeled as
\begin{equation}
    \delta t = M \epsilon + r
\end{equation}
The product of the timing model design matrix $M$ with small offsets in the timing model parameters $\epsilon$ describes the
systematic residuals from subtracting the timing model. $r$ represents the uncorrelated noise in the residuals. \\
[0.15in]
The Narrowband likelihood for the timing residuals is given by 
\begin{equation}
     p\left(\delta t | \epsilon, \phi\right) = \frac{\exp \left(-\frac{1}{2} r^T N^{-1} r\right)}{\sqrt{|2 \pi N|}}
\end{equation}
\begin{equation}
     N_{ij} = (E_{k(i)}^{2} \sigma_i^2 + Q_{k(i)}^{2}) \delta_{ij} 
\end{equation} \\
[0.05in]
where $\phi$ comprises of EFAC $E_{k(i)}$ and EQUAD $Q_{k(i)}$. $\sigma_i$ are the uncertainties in ToAs.\\
[0.15in]
In wideband timing we have an additional likelihood term that includes the DMX priors, 
\begin{equation}
     p\left(\epsilon^{DMX}\;\delta D , E^{DM}\right) = \frac{e^{-\frac{1}{2}\left(\left(\epsilon^{DMX} - \delta D \right)^{T} N^{DM^{-1}}\left(\epsilon^{DMX} - \delta D \right)\right)}}{\sqrt{|2 \pi N^{DM}|}}
\end{equation}
\begin{equation}
     N_{ij}^{DM} = (E_{k(i)}^{DM} \sigma_i^{DM})^2 \delta_{ij} 
\end{equation} \\
[0.05in]
where $E^{DM}$ is the DM EFAC and $\sigma_i^{DM}$ is the DM error. $\epsilon^{DMX}$ represents subset of timing model offsets $\epsilon$ that describe the piece-wise constant DMX model. $\delta D$ is the vector containing difference of DM measurements with respect to the fiducial dm. \\
[0.15in]
The complete wideband timing likelihood is given by the product of both narrowband likelihood and likelihood containing DM priors, \cite{AlamArzoumanian+2021}
\begin{equation} \label{total_likelihood}
    p \left(\epsilon,\phi,E^{DM}|\delta t, \delta D\right) \propto p\left(\delta t | \epsilon, \phi\right) \times p\left(\epsilon^{DMX} | \delta D , E^{DM}\right)\
\end{equation}\\
[0.05in]
This wideband likelihood is implemented in python using \libstempo{}, a python wrapper for \tempotwo{}. We obtained the design matrix $M$ from \libstempo{} by giving the \texttt{par} file and ToAs for a particular pulsar as inputs. This design matrix is then extended to account for $\delta D$ and $\epsilon^{DMX}$. Using the extended design matrix, we estimate the timing residuals and DMX parameters from equation \ref{total_likelihood} using the Generalized Least Squares (GLS) method.



\bsp	
\label{lastpage}
\end{document}